\documentclass[twocolumn]{aastex631}

\usepackage{graphics, graphicx}
\usepackage{subfigure}
\usepackage{amsmath}
\usepackage{threeparttable}
\usepackage[mathlines]{lineno}

\newcommand{\kms}{\hbox{km\,s$^{-1}$}}

\newcommand{\degree}{\hbox{$^{\circ}$}}

\def\OIII{O\,{\sc iii}}

\def\Ha{H{$\rm{\alpha}$}}
\def\Hb{H{$\rm{\beta}$}}

\def\NII{N\,{\sc ii}}
\def\SII{S\,{\sc ii}}
\def\w2026{W2026$+$0716}

\begin{document}

\title{Multi-Component Ionized
 Gas Outflows in a Hot Dust-Obscured Galaxy W2026$+$0716 with Keck/OSIRIS
 \shorttitle{Ionized Gas Outflows in Hot DOG W2026$+$0716}}
 \shortauthors{Liu et al.}

\correspondingauthor{Chao-Wei Tsai}
\email{cwtsai@nao.cas.cn}

\affiliation{National Astronomical Observatories, Chinese Academy of Sciences, 20A Datun Road, Beijing 100101, China}
\affiliation{Institute for Frontiers in Astronomy and Astrophysics, Beijing Normal University,  Beijing 102206, China}
\affiliation{School of Astronomy and Space Science, University of Chinese Academy of Sciences, Beijing 100049, China}

\author[0009-0001-9867-8754]{Chao Liu}
\affiliation{National Astronomical Observatories, Chinese Academy of Sciences, 20A Datun Road, Beijing 100101, China}

\affiliation{School of Astronomy and Space Science, University of Chinese Academy of Sciences, Beijing 100049, China}

\author[0000-0002-9390-9672]{Chao-Wei Tsai}
\affiliation{National Astronomical Observatories, Chinese Academy of Sciences, 20A Datun Road, Beijing 100101, China}
\affiliation{Institute for Frontiers in Astronomy and Astrophysics, Beijing Normal University,  Beijing 102206, China}
\affiliation{School of Astronomy and Space Science, University of Chinese Academy of Sciences, Beijing 100049, China}

\author{Peter R. M. Eisenhardt}
\affiliation{Jet Propulsion Laboratory, California Institute of Technology, Pasadena, CA 91109, USA}

\author[0000-0003-1470-5901]{Hyunsung D. Jun}
\affiliation{Department of Physics, Northwestern College, Orange City, 51041, IA USA}

\author[0000-0003-4007-5771]{Guodong Li}
\affiliation{National Astronomical Observatories, Chinese Academy of Sciences, 20A Datun Road, Beijing 100101, China}
\affiliation{Kavli Institute for Astronomy and Astrophysics, Peking University, Beijing 100871, China}

\author[0000-0001-7808-3756]{Jingwen Wu}
\affiliation{School of Astronomy and Space Science, University of Chinese Academy of Sciences, Beijing 100049, China}
\affiliation{National Astronomical Observatories, Chinese Academy of Sciences, 20A Datun Road, Beijing 100101, China}

\author[0000-0002-9508-3667]{Roberto J. Assef}
\affiliation{Instituto de Estudios Astrof\'isicos, Facultad de Ingenier\'ia y Ciencias, Universidad Diego Portales, Av. Ej\'ercito Libertador 441, Santiago, Chile}

\author[0000-0001-7489-5167]{Andrew W. Blain}
\affiliation{School of Physics and Astronomy, University of Leicester, LE1 7RH Leicester, UK}

\author[0000-0002-2248-6107]{Maren Cosens}
\affiliation{The Observatories, Carnegie Institution for Science, 813 Santa Barbara Street, Pasadena, CA 91101, USA}

\author[0000-0003-0699-6083]{Tanio D\'iaz-Santos}
\affiliation{Institute of Astrophysics, Foundation for Research and Technology-Hellas (FORTH), Heraklion, GR-70013, Greece}
\affiliation{School of Sciences, European University Cyprus, Diogenes Street, Engomi, 1516 Nicosia, Cyprus}

\author[0000-0002-7714-688X]{Román Fernández Aranda}
\affiliation{Department of Physics, University of Crete, 70013, Heraklion, Greece}
\affiliation{Institute of Astrophysics, Foundation for Research and Technology - Hellas (FORTH), Voutes, 70013 Heraklion, Greece}

\author[0000-0003-2478-9723]{Lei Hao}
\affiliation{Shanghai Astronomical Observatory, Chinese Academy of Sciences, Shanghai, China}

\author[0000-0002-9137-7019]{Mai Liao}
\affiliation{National Astronomical Observatories, Chinese Academy of Sciences, 20A Datun Road, Beijing 100101, China}
\affiliation{Universidad Diego Portales, Av Republica 180, Santiago, Reg\'{i}\'{o}n  Metropolitana, Chile}

\author{Shuai Liu}
\affiliation{School of Astronomy and Space Science, University of Chinese Academy of Sciences, Beijing 100049, China}
\affiliation{National Astronomical Observatories, Chinese Academy of Sciences, 20A Datun Road, Beijing 100101, China}

\author[0000-0003-2686-9241]{Daniel Stern}
\affiliation{Jet Propulsion Laboratory, California Institute of Technology, 4800 Oak Grove Drive, Pasadena, CA 91109, USA}

\author[0000-0002-0710-3729]{Andrey Vayner}
\affiliation{IPAC, California Institute of Technology, 1200 E. California Boulevard, Pasadena, CA, USA}

\author[0000-0003-1034-8054]{Shelley Wright}
\affiliation{Center for Astrophysics \& Space Sciences, University of California San Diego, USA}
\affiliation{Department of Physics, University of California San Diego, USA}

\author[0000-0002-4037-3114]{Sherry Yeh}
\affiliation{W. M. Keck Observatory, Waimea, HI 96743, USA}

\begin{abstract}
We present narrowband-filtered integral field unit (IFU) observations of the Hot Dust-Obscured Galaxy (Hot DOG) \textit{WISE} J202615.27+071624.0 (hereafter W2026$+$0716) at redshift $z=2.570$ using Keck/OSIRIS. Our analysis reveals a multi-component ionized gas outflow structure in this heavily obscured AGN host galaxy. Multi-component Gaussian decomposition of the [\OIII] and \Ha\ emission lines uncovers extremely broad and asymmetric profiles, characteristic of AGN-driven outflows. Kinematic mapping shows spatially distinct structures: the [\OIII] and \Ha\ dominated components (with radii of $1.20 \pm 0.56$ kpc) are separated by a projected offset of $\sim 1.1$ kpc and exhibit divergent velocity regimes. The [\OIII] outflow reaches a velocity of 3210 $\pm$ 50 km s$^{-1}$, while the \Ha\ outflow component attains 2310 $\pm$ 840 km s$^{-1}$. Dynamical modeling supports a biconical outflow structure, with [\OIII] and \Ha\ emissions dominating separate cones and significant dust obscuration of the redshifted outflow. Their comparable momentum outflow rates and energy outflow rates suggest a potential physical connection in their driving mechanisms. Spectral energy distribution (SED) analysis reveals anomalous optical/UV excess, attributed to AGN photon scattering by dust or outflowing material, classifying W2026+0716 as a ``Blue Hot DOG.'' The derived outflow timescale ($\sim10^{5}$ yr) aligns with the evolutionary phase of Blue Hot DOGs, suggesting AGN feedback operates persistently during this transitional stage. 

\end{abstract}

\section{Introduction} \label{sec:intro}

Active Galactic Nucleus (AGN) feedback injects
energy and momentum into the surrounding 
interstellar medium (ISM) in the host galaxy through various mechanisms \citep{2010A&A...518L.155F, 2012ARA&A..50..455F, 2015ARA&A..53..115K, 2014A&A...562A..21C}. This feedback process typically impacts the distribution and kinematic properties of gas in galaxies \citep{2012ARA&A..50..455F}, the star formation rate \citep{2005Natur.433..604D, 2016MNRAS.458..816H}, and further affects the evolution of the host galaxy \citep{2013ARA&A..51..511K, 2014ARA&A..52..589H}. AGN-driven outflows play a crucial role in the feedback process \citep{2014ARA&A..52..529Y, 2015ARA&A..53..115K, 2018NatAs...2..176C}. The energy or momentum of these AGN-powered outflows \citep{2012MNRAS.425..605F, 2014MNRAS.444.2355C} can eject material from the nuclear region of galaxies at high velocities \citep{2011ApJ...729L..27R, 2015A&A...583A..99F, 2017A&A...601A.143F}. 

AGN feedback is expected during the short “blowout” phase of an obscured AGN in merging driven galaxy evolution. During this phase, AGN outflows clear the surrounding dust and gas, revealing bright, unobscured quasars, which may further deplete the cold gas available for star formation. This process drives the transition of merging galaxies from an early dust-obscured, star-forming, and rapidly accreting AGN phase to a later phase characterized by suppressed star formation and AGN dominance \citep[e.g.,][]{1988ApJ...325...74S, 2016MNRAS.458..816H}. Within this framework, dust-reddened, obscured AGNs offer a key sample for investigating this rapid transitional evolution process. These obscured AGN systems often show active star formation \citep[e.g.,][]{2001ApJ...555..719C, 2008ApJ...674...80U, 2015ApJ...804...27A, 2021A&A...649A.102C, 2022ApJ...934..119G,
2024ApJ...964...95S}, suggesting that the AGN energy output has begun clearing the surrounding medium but has not yet fully halted star formation activity. Dust-reddened and obscured AGNs are thus prime candidates for studying how AGN feedback regulates galaxy evolution.

Hot Dust-Obscured Galaxies (Hot DOGs)
are a population of highly obscured, hyper-luminous galaxies \citep{2012ApJ...755..173E, 2012ApJ...756...96W}
discovered by the \textit{WISE} survey \citep{2010AJ....140.1868W}. They represent the most luminous obscured galaxies in the universe with luminosities exceeding $10^{13} 
L_\odot$ (over 10\% of them surpassing $10^{14} 
L_\odot$)
\citep{2015ApJ...805...90T}, spanning a wide redshift range of $ z \sim 1$--$4.6$ \citep[][; Eisenhardt et al. in prep.]{2012ApJ...755..173E, 2024ApJ...971...40L} with few sources at $z \sim 0.365-0.465$ \citep{2023ApJ...958..162L} and are commonly found in high-density environments \citep[][]{2014MNRAS.443..146J, 2015ApJ...804...27A, 2018Sci...362.1034D, 2022NatCo..13.4574G, 2022ApJ...935...80L, 2023A&A...677A..54Z, 2024arXiv241204436Z}. Multi-wavelength observations with space facilities and ground-based telescopes show a strong mid-infrared component in their spectral energy distributions (SEDs), suggesting higher dust temperatures \cite[$\sim$ 450\,K or higher,][]{2015ApJ...805...90T} compared to typical dust-obscured galaxies \citep{2003ASPC..289..247B, 2009ApJ...693..750B, 2014PhR...541...45C}. They are indeed rare, with a surface number density of $0.032 \pm 0.004 \, \mathrm{deg}^{-2}$ (corresponding to one object per $31 \pm 4 \, \mathrm{deg}^{-2}$), yet their number density is on par with that of equally luminous, unobscured quasars at the brightest end of the luminosity function \citep{2015ApJ...804...27A}. The extreme luminosities of Hot DOGs are powered by accretion onto a central supermassive black hole (SMBH) heavily enshrouded in gas and dust \citep{2012ApJ...755..173E, 2015ApJ...804...27A, 2015ApJ...805...90T}, making them difficult to detect in the X-ray band due to the high Compton thickness \citep{2014ApJ...794..102S, 2015A&A...574L...9P, 2016ApJ...819..111A, 2018MNRAS.474.4528V, 2020ApJ...897..112A}. The black hole masses have been estimated at around $10^{9}\,\mathrm{M}_{\sun}$ with an Eddington ratio close to unity \citep{2018ApJ...852...96W,2020ApJ...905...16F,2024ApJ...971...40L} and occasionally surpass the Eddington limit \citep{2018ApJ...868...15T, 2024ApJ...971...40L}. The extreme luminosities ($>10^{13} L_{\odot}$) of these highly obscured AGNs within typical galaxies \citep{2015ApJ...804...27A, 2024ApJ...971...40L} suggests that Hot DOGs are experiencing a critical transitional phase in galaxy and AGN evolution, characterized by high luminosity, high accretion, and intense feedback.

Blue-excess Hot Dust-obscured Galaxy (BHD), first identified as a distinct subpopulation of Hot DOGs by \cite{2016ApJ...819..111A}, exhibit significant blue/UV excess emission in their rest-frame SEDs. Unlike typical Hot DOGs where the blue/UV continuum can be sufficiently modeled with star formation components, BHDs' SEDs require incorporation of an AGN component dominating $>50\%$ of the bolometric luminosity blueward of 1 \micron\, to obtain physically consistent SED solutions \citep{2024ApJ...971...40L}. Polarimetric observations together with radiative transfer modeling reveal that the blue-excess emission predominantly originates from scattered light emerging through partial obscuration of a centrally-located, heavily dust-enshrouded AGN \citep{2020ApJ...897..112A, 2022ApJ...934..101A}.

Evidence of outflows in Hot DOGs has been observed across multiple wavelengths. Spectroscopic signatures of kpc-scale ionized gas outflows, based on the broad and blue-shifted [\OIII] emission lines \citep[][]{2020ApJ...905...16F, 2020ApJ...888..110J, 2024arXiv241202862V} and polarimetric imaging (\citealt[Assef et al. in prep.]{2022ApJ...934..101A}), suggest that these objects commonly host high-kinetic-energy outflows, indicating strong AGN feedback. Recent studies \citep{2024MNRAS.534..978M} have expanded the sample of outflows and provided the first evidence for molecular gas outflows. In some Hot DOGs with significant outflow features, multi-wavelength SEDs display an additional emission component potentially linked to the outflow or AGN photon scattering by dust (\citealt{2022ApJ...934..101A, 2024ApJ...971...40L}, Assef et al. 2024 in prep.). However, these studies on Hot DOG outflows are primarily based on 1-D spectroscopy without spatial information. High spatial resolution integral field spectroscopic observation data can directly reveal the outflow gas morphology, kinematic properties, and its interactions with the ISM of the host galaxy in detail.

In this study, we conducted near-infrared IFU observations of \textit{WISE} J202615.27$+$071624.0 (hereafter W2026$+$0716), at $z$ = 2.570, as spectroscopically measured by \cite{2020ApJ...888..110J}. The \textit{HST} F160W image shows its compact and mildly disturbed morphology \citep{2016ApJ...822L..32F}. This system is identified as a BHD with ionized gas outflow features in its 1-D optical spectrum. Our analysis and discussions focus on the spatial discrepancies between the broad [\OIII] and \Ha\ emission lines. This paper is organized as following: Section \ref{sec:Observation and DATA Deduction} details our observational methods and data processing; Section \ref{sec:Data Analysis} describes our data fitting and analysis techniques; Section \ref{sec:Results} presents the imaging analysis results and kinematic calculations; Section \ref{Discussion} provides a physical interpretation and discussion of our observational and analytical findings; and Section \ref{sec:Conclusion} concludes with a summary of the study. This study assumes a flat \(\Lambda\)CDM cosmology, with \( H_0 = 70 \, \mathrm{km \, s^{-1} \, Mpc^{-1}} \) and \( \Omega_M = 0.3 \). At $z=2.570$, 1\arcsec\ corresponds to 8.02 kpc.

\section{Observation and DATA Reduction}\label{sec:Observation and DATA Deduction}
\subsection{Observations}\label{subsec:Observation}
We conducted near-infrared IFU observations of the target W2026+0716 (RA = 20h 26m 15.27s, Dec = +07$\degree$ 16\arcmin\ 24\farcs02) using OSIRIS \citep{2006SPIE.6269E..1AL, 2014PASP..126..250M} on the Keck I Telescope with Laser Guide Star Adaptive Optics (LGS-AO) system on Oct 7th and 8th, 2019 UTC. We used the 35 mas pupils, incorporating both the \textit{$Hn5$} and \textit{$Kn5$} filters. In this setup, a field of view of 1.12$^{\prime \prime}$$\times$2.24$^{\prime \prime}$ was attained, along with a sampling of 0.035$^{\prime \prime}$. The observations covered the wavelength ranges 1.721-1.808 \micron\ and 2.292-2.408 \micron\ with spectral resolutions of $R \approx 3800$. The angular resolution was calibrated through standard star observations, obtaining the full width at half maximum (FWHM) of the point spread function  $\sim 0.092$$^{\prime \prime}$ and $\sim 0.098$$^{\prime \prime}$ in the $Hn5$ and \textit{$Kn5$} bands, respectively. We obtained three 900-second exposures for each band with dithering of 0.175$^{\prime \prime}$ between consecutive observations, resulting in a total integration time of 45 minutes on W2026+0716. Similar dithering approach was applied to the observations of the standard stars with a typical integration time of 9 seconds per exposure.

\subsection{Data Reduction}\label{subsec:DATA Reduction}
The raw data obtained from Keck/OSIRIS were processed into data cubes using the Keck/OSIRIS Data Reduction Pipeline \citep[DRP,][]{2017ascl.soft10021L, 2019AJ....157...75L}. We applied standard processing procedures to all observational data using the OSIRIS DRP including dark subtraction, channel-level adjustments, crosstalk removal, glitch identification, cosmic ray cleaning, and data cube assembly. The detailed descriptions of these procedures can be seen in the OSIRIS User Manual\footnote{\url{https://www2.keck.hawaii.edu/inst/osiris/OSIRIS_Manual.pdf}}. Subsequently, we employed the OH line suppression scaling sky subtraction algorithm based on \citet{2007MNRAS.375.1099D}, which effectively removed OH sky emission lines from the observational data.

We performed atmospheric telluric correction and flux calibration for both \textit{$Hn5$} band and \textit{$Kn5$} band observations
using the standard stars. Then we performed spectral spike cleaning by removing anomalous signals in each data cube. These spikes were identified through a localized $5\sigma$ thresholding process: Any individual spaxel showing flux values exceeding $5\sigma$ above the background level within its immediate $3\times3$ spaxel neighborhood was flagged as an anomaly. These outliers were then replaced by the median value computed from the 8 surrounding spaxels (excluding the central spike itself), ensuring the preservation of local background continuity while eliminating uncorrelated noise events. The number of anomalous spikes is $<5\%$ of the total 3-D data.

For each exposure frame, we selected the median spectra from several source-free regions as a sky background and subtracted this background from each pixel's spectral data to obtain the clean datacube. Finally, we combined multiple frames of data in the same band using a weighted average method to produce the final datacube. To optimize the signal-to-noise ratio, we set the weight of each velocity channel as the ratio of the observational data to the observational error. We spatially aligned the $Hn5$ band and $Kn5$ band datacubes using the peaks of their continuum emissions before conducting data analysis. More detailed discussions about the spatial alignment are included in Section \ref{subsec:Spectrum Fitting}.

\section{Data Analysis} \label{sec:Data Analysis}
\subsection{SED Fitting } \label{subsec:SED Fitting}

\begin{figure*}[ht!]
\centering
\includegraphics[width=\textwidth]{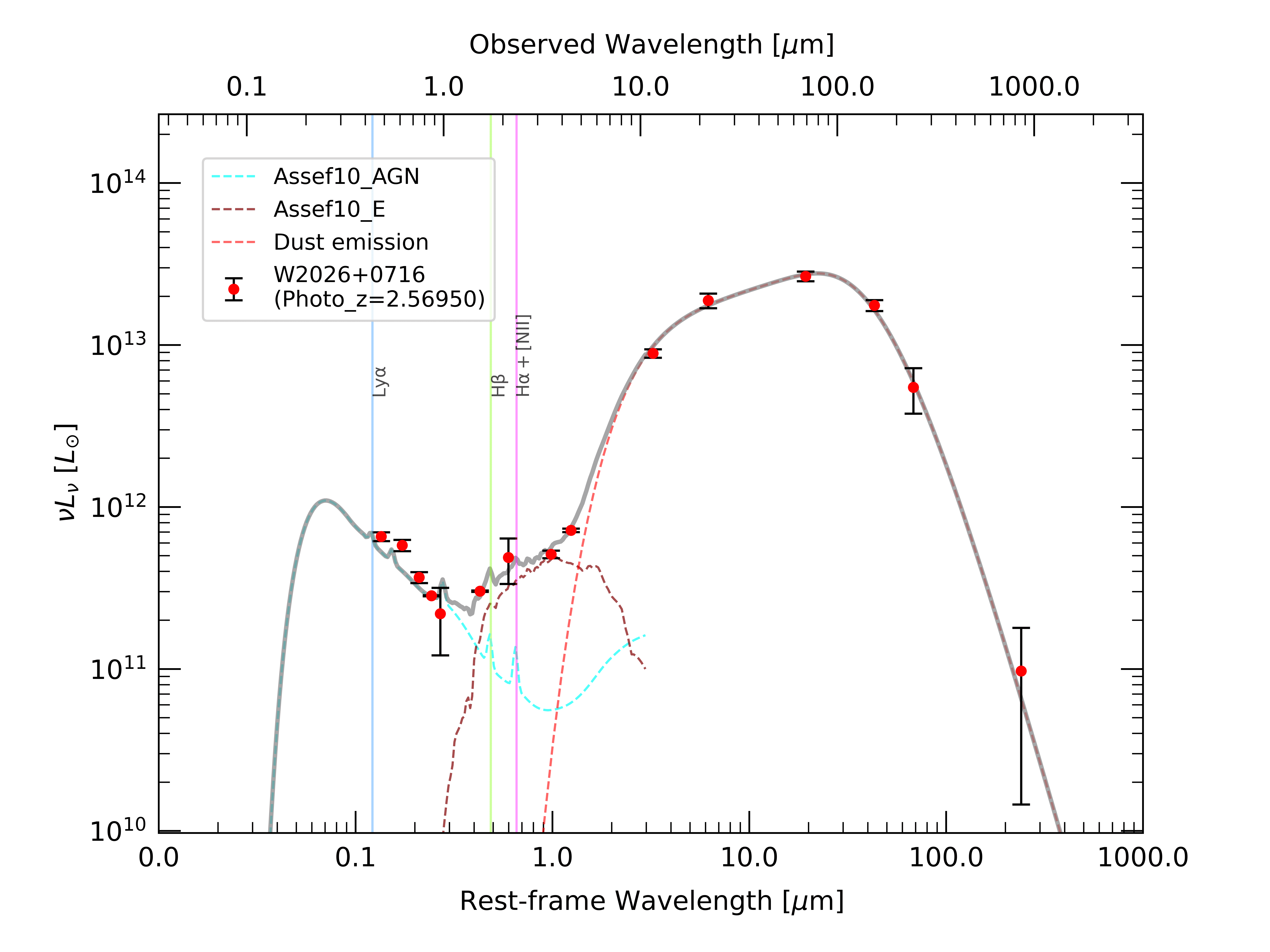}
\caption{\textbf{SED fitting results for 
W2026$+$0716
} The red points with error bars represent the multi-band photometric values from Table \ref{tab:table1}, and the gray curve is the best fit curve obtained through the MCMC algorithm. The cyan curve represents a low-extinction AGN model, indicating contributions from unobscured AGN, while the brown fitting line represents a model for elliptical galaxies, highlighting the contribution of older stellar components. These models are adopted from \citet{2010ApJ...713..970A}.
The dark orange line represents the hot dust emission using the SPL model, from Tsai et al. (in preparation).}
\label{fig:SEDFitting}
\end{figure*}

We performed SED fitting of W2026+0716 using photometry from ultraviolet to sub\-millimeter wavelengths. Our dataset \footnote{The HST data used in this paper can be found in
MAST: \dataset[10.17909/y5pw-cx63]{http://dx.doi.org/10.17909/y5pw-cx63}.  The data from WISE, Spitzer, and Herschel can be found in IRSA: \citet{AllWISE2019}, \citet{Spitzer2020}, and \citet{Herschel2021}, respectively.}encompasses observations from multiple facilities including SDSS, Pan-STARRS1, \textit{HST}, Hale 200 inch, \textit{Spitzer} IRAC, \textit{WISE}, \textit{Herschel} PACS and SPIRE, as well as JCMT SCUBA-2, see Table \ref{tab:table1}.

The UV-to-MIR SED modeling was conducted following the methodology developed by \citet{2010ApJ...713..970A, 2016ApJ...819..111A}. As shown in Figure \ref{fig:SEDFitting}, for the short-wavelength regime ($\lambda < 3\mu m$) in the rest frame, we employed a linear combination of three templates: one AGN template with variable extinction and two galaxy templates representing stellar populations of different ages. This approach enabled us to simultaneously account for contributions from both AGN and stellar radiation. For the long-wavelength regime ($\lambda > 3\mu m$), we implemented a single power-law (SPL) dust distribution model to capture the thermal radiation from the dust-obscured AGN (Tsai et al. in prep). The fitting procedure utilized a Markov Chain Monte Carlo (MCMC) method with five free parameters, including the amplitudes of individual templates, AGN extinction level, and the power-law index of the SPL model.

Based on our best-fit model, we derived several key physical properties of \w2026. The object exhibits a total luminosity of $6.7 \times 10^{13} L_\odot$, indicating an extremely luminous source. The model reveals substantial hot dust content with an estimated total mass of $1.4 \times 10^{7} \mathrm{M}_{\odot}$. Analysis of the stellar component suggests a star formation rate of $\sim130\,\mathrm{M}_\odot/yr$
and a median stellar mass of $8.0 \times 10^{11} \mathrm{M}_{\odot}$. We note that
our analysis reveals significant excess blue emission in the optical bands, likely originating from scattered AGN light. This characteristic qualifies \w2026
as a ``Blue Hot DOG" (BHD), suggesting that a portion of the central AGN radiation scatters off the obscuring dust layer. As a result, the stellar mass and star formation rate estimates from the SED modeling might be affected by the leaked AGN emission.

\begin{deluxetable}{lCCC} 
\tablecaption{Multiband Photometry of Hot DOG W2026+0716 \label{tab:table1}}
\tablewidth{0pt}
\tablehead{
\colhead{Band} & \colhead{Wavelength$^{\rm a}$} & \colhead{Flux} & \colhead{Ref.$^{\rm b}$} \\
\colhead{} & \colhead{($\mu$m)} & \colhead{($\mu$Jy)}
}
\startdata
PS1.g  & 0.48 & 5.2 \pm 0.5 & \textbf{\textit{C16}};\textbf{\textit{F20}}\\
PS1.r  & 0.62 & 6.5 \pm 0.7 & ... \\
PS1.i  & 0.75 & 5.5 \pm 0.5 & ... \\
PS1.z  & 0.87 & 5.1 \pm 0.1 & ... \\
PS1.y  & 0.96 & 4.5 \pm 2.3 & ... \\
\textit{HST}.WFC3.F160W  & 1.53 & 10.5 \pm 0.1 & \textbf{\textit{L25}}\\
Hale 200 inch.WIRC  & 2.13 & 24.2 \pm 7.8 & \textbf{\textit{A15}}\\
\textit{Spitzer}.IRAC.I1 & 3.54 & 42.5 \pm 2.3 & \textbf{\textit{G12}}\\
\textit{Spitzer}.IRAC.I2 & 4.44 & 75.9 \pm 1.9 & ... \\
\textit{WISE}.W3 & 11.56 & 2.5 \pm 0.2 \times 10^{3} & \textbf{\textit{W10}};\textbf{\textit{C13}}\\
\textit{WISE}.W4 & 22.09 & 10.0 \pm 1.0 \times 10^{3} & ... \\
\textit{Herschel}.PACS.blue & 68.92 & 44.0 \pm 3.0 \times 10^{3} & \textbf{\textit{T25}} \\
\textit{Herschel}.PACS.red & 154 & 65 \pm 5 \times 10^{3} & ... \\
\textit{Herschel}.SPIRE.PSW & 243 & 32 \pm 10 \times 10^{3} & ... \\
JCMT.SCUBA2.850 & 858 & < 5.1 \times 10^{3} & \textbf{\textit{J14}}\\
\enddata
\tablecomments{
$^{\rm a}$\,Effective wavelength, adopted from SVO Filter Profile Service \citep{2012ivoa.rept.1015R,2020sea..confE.182R}. 
$^{\rm b}$\,The reference code: 
\textbf{\textit{A15}}: \cite{2015ApJ...804...27A}; 
\textbf{\textit{C13}}: \cite{2014yCat.2328....0C};
\textbf{\textit{C16}}: \cite{2016arXiv161205560C};
\textbf{\textit{F14}}: \cite{2014MNRAS.443..146J}; 
\textbf{\textit{F20}}: \cite{2020ApJS..251....7F}; 
\textbf{\textit{G12}}: \cite{2012AJ....144..148G}; 
\textbf{\textit{L25}}: This work, using basic aperture photometry.;
\textbf{\textit{T25}}: Tsai et al. in prep.; 
\textbf{\textit{W10}}: \cite{2010AJ....140.1868W}.
}
\end{deluxetable}

\subsection{Spectrum Fitting } \label{subsec:Spectrum Fitting}

\begin{figure*}[ht!]
\centering
\includegraphics[width=\textwidth]{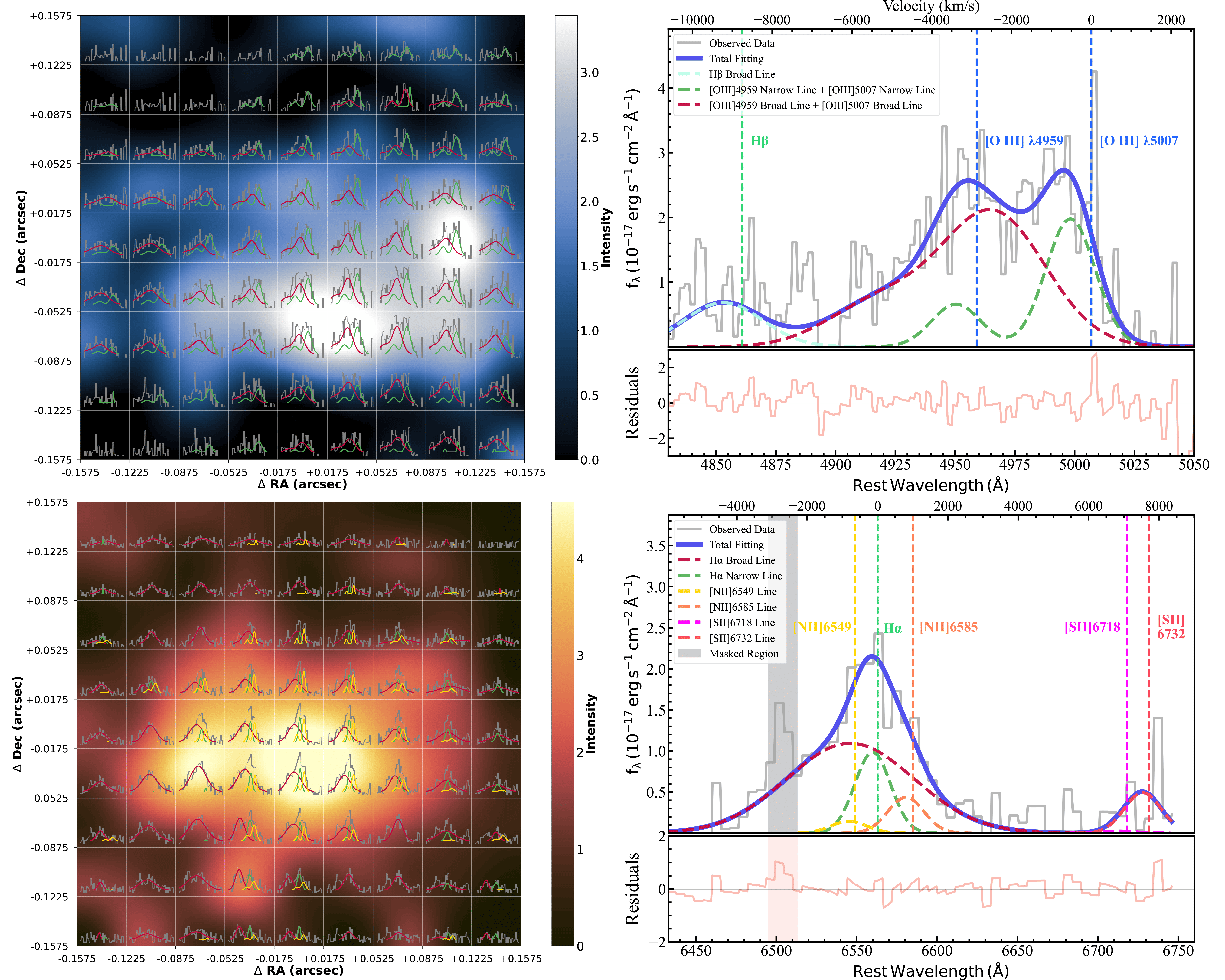}
\caption{
\textbf{Spectral fitting results for W2026+0716} 
The left panels represent the integrated flux background of [\OIII] (\textit{top}) and \Ha\ (\textit{bottom}) overlaid with the spatially resolved, best-fit emission line spectral models of outflow gas.
Each integrated unit represents the fitting result of the data
corresponding to a single pixel.
The background images have been resampled for clear presentation.
In the [\OIII]
panel, the green and red fitting components
represent the narrow and broad components of the [\OIII]$\lambda$4959\AA/5007\AA\ doublet emission, respectively.
In the \Ha\
panel, the green and red fitting components, 
represent the narrow and broad components of the \Ha\
emission line, 
respectively, while the yellow fitting component represents the [\NII]$\lambda$6549\AA/6585\AA\ doublet emission. 
The right panels show the 1-D fitting results of the integrated flux over the emission line windows in 
\textit{$Hn5$} band (\textit{top}) and \textit{$Kn5$} band (\textit{bottom}). 
The green and red fitting components correspond to the broad and narrow components of the emission lines within those bands. There are 32 spectral channels in the \textit{$Hn5$} band that are masked (shaded regions) due to contamination from OH sky emission lines.
}
\label{fig:SpectrumFitting}
\end{figure*}

We modeled the emission lines with a combination of narrow and broad components using PyQSOfit \citep{2018ascl.soft09008G, 2019ApJS..241...34S}, where the
narrow component represents Gaussian at
the systemic velocity with a dispersion $ < 1500$ \kms
and the broad component characterizes the high-velocity outflow with a larger dispersion. [\OIII]$\lambda$4959\AA, [\OIII]$\lambda$5007\AA, and \Ha\
include both broad and narrow components. For other weaker emission lines, we employed a single Gaussian component for only the narrow component. All the spectroscopic data are corrected for Galactic extinction before multi-Gaussian modeling.

For all emission lines within the same spectral band, components originating from the same physical region share fixed line centers and full width at half maximum. Additionally, for doublet lines, their relative flux ratios are constrained during the spectral fitting process. The intensity ratio of [\OIII] and [\NII] doublets was fixed at $1:3$ \citep[according to][]{2000MNRAS.312..813S}. However, the [\SII] doublet ratio was not fixed because of its dependence on the electron density of the ionized gas \citep{2016ApJ...816...23S}. Given the source's weak continuum as a typical characteristic of Hot DOGs, we performed a simple linear fit for the continuum spectrum to prevent the overfitting issues.

For individual spaxel fitting, the spectral data for each pixel were convolved with a Gaussian kernel of an FWHM of 2.5 pixels over a $5\times5$ pixel window. The combined effect of the PSF and Gaussian smoothing yielded an actual spatial resolution of 0\farcs139, which corresponds to a spatial scale of 1.14 kpc. Additionally, we employed PyQSOFit with an “nsmooth” value of 8 to mitigate the fitting instability. Figure \ref{fig:SpectrumFitting} represents the results of the spectral fitting on the \Ha\ and [\OIII] lines. 

\begin{figure}[htbp]
\centering
\includegraphics[width=\columnwidth]{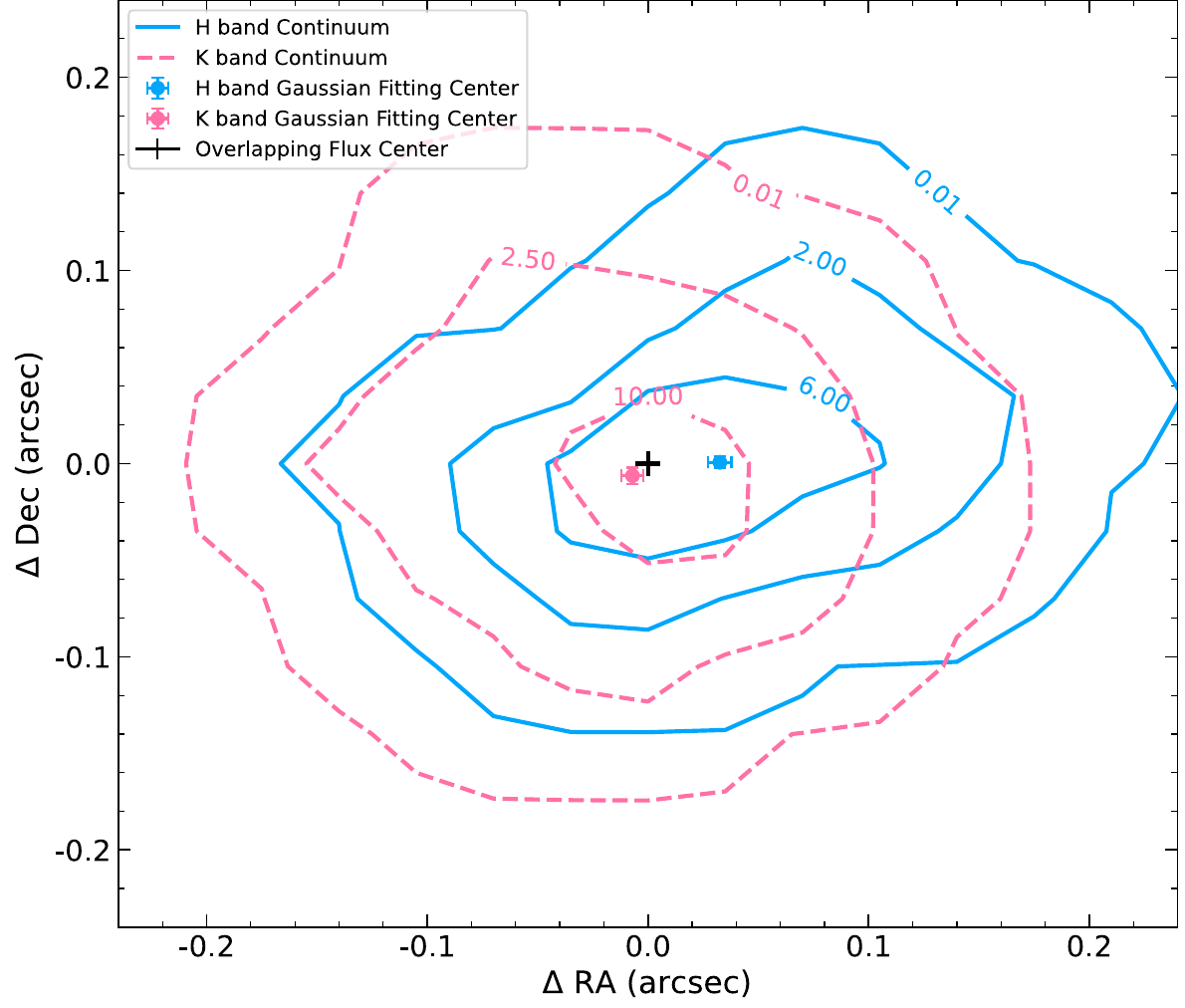}
\caption{
\textbf{Contours of the continuum peaks of \w2026\ in \textit{$Hn5$} and \textit{$Kn5$} bands}. The best-fit Gaussian peaks in two bands show an offset of 0\farcs046 based on the relative astrometry using the laser guide stars. This small offset can be attributed to the different optical paths. We use the overlapping flux peak (marked by black crosses) as the common spatial reference center between the two-band datacubes.
}
\label{fig:ContinuumPeaks}
\end{figure}

\section{Results} \label{sec:Results}

\subsection{Kinematics Maps}\label{subsec:Kinematics Map}

\begin{figure*}[htbp!]
\centering
\includegraphics[width=\textwidth]{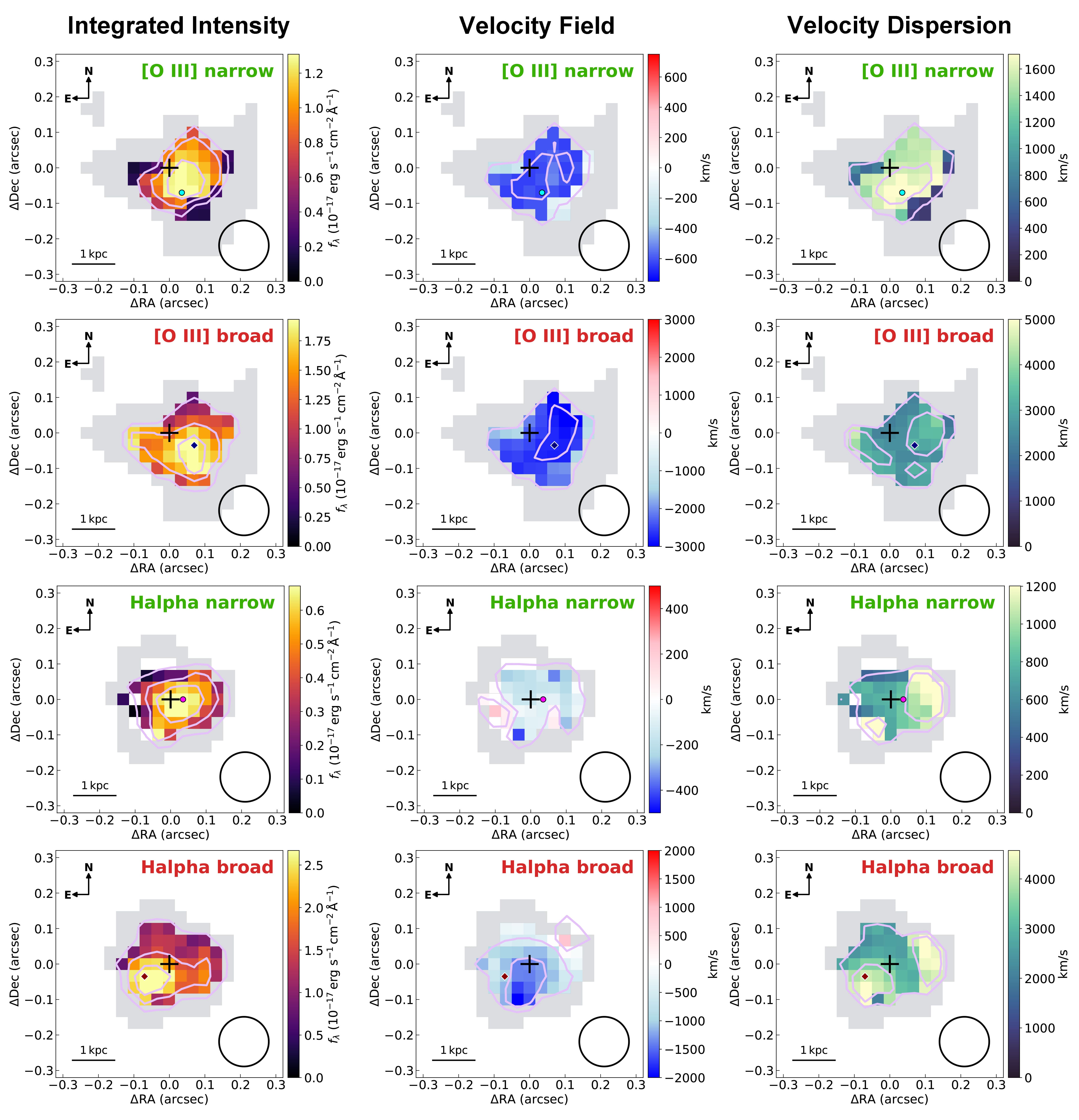}
\caption{\textbf{Kinematics maps of the fitting results for different emission components across spatial locations.} Columns from left to right respectively represent the flux, line center velocity shift, and velocity dispersion of each emission component. Our map was generated without flux weighting, relying instead on the direct model fit. If flux weighting were applied, the large uncertainties could introduce significant biases, particularly in datasets with low signal-to-noise ratios.
Rows from top to bottom are sequentially \Ha\ narrow line, \Ha\ broad line, 
[\OIII]
narrow line, [\OIII]
broad line. Each pixel in the image is the result of convolution over a $5 \times 5$
window multiplied by a Gaussian kernel with a full width at half a maximum of 2.5 pixels.  The contours represent distinct values in each panel, potentially highlighting different structures. 
The black cross at the center marks the position of the AGN, corresponding to the highest flux value within the continuum contours shown in Figure \ref{fig:ContinuumPeaks}. The gray regions represent diffuse gas, and points of different colors indicate the peak flux locations of the corresponding components. Directions, scale, and size of resolution element are annotated in the image, with the resolution for the \textit{$Hn5$} and \textit{$Kn5$} bands being 0\farcs139 and 0\farcs144, respectively.}
\label{fig:MomentMap}
\end{figure*}

We spatially aligned the [\OIII] and \Ha\ datacubes using continuum emission from the \textit{$Hn5$} band and \textit{$Kn5$} bands (as shown in Figure \ref{fig:ContinuumPeaks}). A 2-D Gaussian fit to the continuum contours in both bands yielded FWHMs of $0\farcs175 \pm 0\farcs012$ (major axis) and $0\farcs098 \pm 0\farcs007$ (minor axis) in the \textit{$Hn5$} band, and $0\farcs137 \pm 0\farcs012$ (major axis) and $0\farcs175 \pm 0\farcs012$ (minor axis) in the \textit{$Kn5$} band. The estimated spatial alignment uncertainty between [\OIII] and \Ha\ maps is $0\farcs04$. With the datacubes properly aligned, we compiled the spectral fitting results for each pixel into maps, allowing us to examine the spatial distribution of various spectral line components. We constructed the spatial distribution of flux intensity, velocity offset, and velocity dispersion of the [\OIII] and \Ha\ lines in Figure \ref{fig:MomentMap}, to study the kinematics of the ionized gas in \w2026. 

Examining the flux intensity maps, we find that the narrow components of both emission lines are concentrated near the AGN (marked with black cross in the Figure \ref{fig:MomentMap}) with slight offsets. The broad components, representing outflows, show more pronounced spatial offsets from the continuum peak, which is presumably the location of the obscured AGN and the assumed galaxy center. These significant
offsets exceed half a size of resolution element and the spatial alignment uncertainty of $0\farcs04$. The flux peaks
of both [\OIII] and \Ha\ outflows are offset with respect to the galaxy center by similar amounts but in quite different directions. The peak of [\OIII]
outflow's flux is offset 0\farcs08 to the southwest from the continuum emission peak, while the \Ha\ outflow's flux peak is offset 0\farcs08 toward the southeast.

Furthermore, the extent of [\OIII] and \Ha\ diffuse gas regions also shows a discrepancy. The spatially [\OIII] diffuse gas exhibits a more extensive region, which is shown in gray in Figure \ref{fig:MomentMap}, and exhibits less gas emission in the southeast where the \Ha\ outflow emission dominates.

The velocity fields of the narrow and broad components for both [\OIII] and \Ha\
emission lines show overall blueshifted velocities without significant gradients. The narrow component of \Ha\
 exhibits blueshifted velocities mostly within 400 \kms, with some regions showing slight redshift. The [\OIII] narrow component shows slightly higher blueshifted velocities, mostly within 600 \kms. For the broad components, the \Ha\
outflow has an overall blueshift exceeding 500 \kms, with a peak blueshifted velocity over 1200 \kms
in the southeast, corresponding to its flux peak. The [\OIII] outflow shows even higher overall blueshifted velocities exceeding 1000 \kms, with a peak of over 2600 \kms in the southwest, aligning with its flux center. Notably, the \Ha\
outflow exhibits weak blueshift or even redshift in the northwest at a 0\farcs1 offset from the continuum peak, though the redshifted outflow is detected in only two pixels. These features may relate to irregularity in the 2-D structure observed in the velocity dispersion map. The high blueshifted velocities of broad components of both lines coupled with their spatially corresponding high dispersion regions, provide strong evidence for outflows.

The velocity dispersion maps of the [\OIII] and \Ha\ broad components reveal intriguing kinematic
features. The broad-line components of both [\OIII] and \Ha\  exhibit velocity dispersions exceeding 1500 \kms, which suggests significant turbulence within the ionized gas outflows of \w2026. Broad components of both lines display two high-velocity dispersion regions on the east and west sides of the continuum emission peak. However, the widths of [\OIII] broad components in the high dispersion regions exceed 2700 \kms, while those of the \Ha\ broad components surpass 3500 \kms\ in the corresponding positions.

Such broad line widths could result from the superposition of different velocity components along the line of sight or from turbulent regions created by outflow interactions with the surrounding interstellar medium. Either scenario implies the presence of multi-velocity component regions in the ionized gas, which is
consistently observed in emission lines from two observed wavebands of \w2026\ system. 
Combining these findings with our analysis of flux intensity and velocity offset distributions, they suggest the presence of high-velocity ionized gas outflows in Hot DOG \w2026.

\subsection{Dynamics of Outflows}\label{subsec:Outflows Energetics Calculation}

We obtained fundamental kinematic parameters of the ionized gas to calculate the dynamical characteristics of the outflows. These pieces of information allow us to investigate their origins, driving mechanisms, and evolutionary scenarios.

The outflow radius indicates a representative maximum spatial extent of outflow, which is crucial for calculating other dynamical parameters. In previous studies, the outflow radius for Hot DOGs was often assumed to be 3 kpc \citep{2020ApJ...905...16F, 2020ApJ...888..110J, 2024MNRAS.534..978M}, which serves as a conservative estimate. We instead estimate a spatially resolved outflow radius utilizing the LGS-AO guided OSIRIS observations. However, the overall blueshifted outflows indicate a projection of these outflows along our line of sight. While existing information is insufficient to determine the exact projection angle, the overall blueshift in the velocity field indicates that this angle is relatively small. Considering that the outflow axis being exactly aligned with the line of sight is a low-probability event, we adopted a projection angle $i$ of $30\degree$ \citep{2016ApJ...828...97B}. Thus, we applied the following corrections to the outflow radius and velocity using $i = 30\degree$:

\begin{equation}
R_{\rm out} =  \frac{R_{\rm out, observed}}{\sin i},
\end{equation}
\begin{equation}
\nu_{\rm out} = \frac{\sqrt{\Delta \nu_{\,\rm broad}^2 + \sigma_{\,\rm broad}^2}}{\cos i } 
\end{equation}

A 30$\degree$ inclination angle may be a conservative estimate, with the actual angle potentially larger or smaller. This inclination angle typically does not exceed 45$\degree$, or the redshifted outflow would likely be observed. If the outer opening angle of the outflow lies between 30$\degree$ and 45$\degree$, we may overestimate the outflow radius and underestimate its velocity. As a result, the observed outflow could be distributed more confined in a smaller region with higher velocity. This compact, high-velocity outflow could represent a younger outflow in an early formation phase. On the other hand, if the opening angle is between 0$\degree$ and 30$\degree$, the outflow could be denser and in a more mature stage, possibly interacting with the interstellar medium (ISM).

In the kinematic maps (Figure \ref{fig:MomentMap}), the correspondence between the outflow’s flux peaks and its high-velocity dispersion region allows us to distinguish the outflow radius in the plane of the sky. Despite the a spatial offset of $\sim 1/3$ size of resolution element between the two, the displacement of flux centroids used to determine the outflow radius is a relatively conservative and reliable measurement. Therefore, we consider the $0\farcs075$ offset from the galactic center of this region as the outflow radius. After deprojection correction, the outflow radius $R_{\rm out}$ is estimated as $0\farcs15$, which corresponds to 1.2 kpc for \w2026.

\begin{deluxetable*}{llccl}[!htbp]
\tablecaption{
Dynamics Parameters of Hot DOG \w2026
\label{tab:table2}}
\tablewidth{0pt}
\tablehead{
\colhead{Parameters} & \colhead{Meaning} & \colhead{[\OIII] Outflow} & \colhead{\Ha\ Outflow} & \colhead{Unit}
}
\startdata
$L_\mathrm{out}$  & Outflow Luminosity  & $7.82 \pm 0.48$&  $5.27 \pm 0.14$ & $10^{43}$ erg s$^{-1}$ \\
$v_\mathrm{br}^{\rm obs}$ & Observed Velocity of Outflow  & {$-2450\pm 30$} & {$-850\pm 590$} & km s$^{-1}$\\
$v_\mathrm{out}$   & Projection Corrected Velocity of Outflow  & $3210\pm 50$ & $2310\pm 840$ & km s$^{-1}$ \\
$\sigma_\mathrm{out}$    & Velocity Dispersion of Outflow  & $1310\pm 20$ & $1810\pm 370$ & km s$^{-1}$ \\
$R_\mathrm{out}$   & Radius of Outflow  & $1.20 \pm 0.56$ & $ 1.20 \pm 0.56$ & kpc \\
$M_\mathrm{out}$   & Mass of Outflows  &
$6.3 \pm 0.4$ & $16 \pm 4$ & $10^{7} \mathrm{M}_{\sun}$ \\
$\dot{M}_\mathrm{out}$ & Mass Outflow Rate & $510 \pm 240$ & $930 \pm 600$ & $M_{\odot}$/yr \\
\enddata
\tablecomments{
Dynamical
parameters for the 
[\OIII] and \Ha\
outflows in the 
Hot DOG
W2026$+$0716. 
The electron density ($N_e$) is assumed to be 500 cm$^{-3}$ for both outflows in the calculations. 
See Section \ref{subsec:Outflows Energetics Calculation} for details.
}
\end{deluxetable*}

Electron density constitutes a crucial parameter in determining outflow masses. The [\SII] emission doublet, widely recognized for its electron density sensitivity \citep{2016ApJ...816...23S}, serves as the standard diagnostic. However, in our observations of W2026+0716, the [\SII] doublet was only marginally detected, permitting single-component Gaussian fitting and precluding robust electron density measurements. Therefore, we adopted an assumption-based approach to estimate the electron density.

AGN-driven outflows typically exhibit electron densities in the range $N_{e} = 100-1000$ cm$^{-3}$ \citep{2011ApJ...732....9G, 2014MNRAS.441.3306H, 2016ApJ...833..171K}. Previous studies of Hot DOG outflows adopted $N_{e} = 300$ cm$^{-3}$ based on an assumed characteristic radius of 3 kpc \citep{2020ApJ...905...16F, 2020ApJ...888..110J}. For W2026+0716, however, the measured outflow radius of 1.2 kpc implies that using $N_{e} = 300$ cm$^{-3}$ would overestimate the mass outflow rate due to the $N_{e}$ scaling dependence in mass calculations. We therefore conservatively assume $N_{e} = 500$ cm$^{-3}$ to mitigate potential overestimation of feedback effects.

The outflow masses are calculated following the formalism of \cite{2015A&A...580A.102C} and \cite{2020A&A...644A..54S},

\begin{equation}
\frac{M_{\textrm{$\mathrm{[O \, III]}$, out}}}{\mathrm{M}_{\odot}} = 0.4 \times 10^8 \left(\frac{L_{\textrm{$\mathrm{[O \, III]}$}}}{10^{43} \textrm{ erg s}^{-1}}\right)\left(\frac{N_e}{100 \textrm{ cm}^{-3}}\right)^{-1}
\end{equation}
and
\begin{equation}
\frac{M_{H\alpha, \text{ out} }}{\mathrm{M}_{\odot}} = 
1.5 \times 10^8 \left(\frac{L_{H\alpha}}{10^{43} \text{ erg s}^{-1}}\right)\left(\frac{N_e}{100 \text{ cm}^{-3}}\right)^{-1}
\end{equation}

We find the outflow masses reach $6.3 \times 10^{7} \mathrm{M}_{\sun}$ and $1.6 \times 10^{8} \mathrm{M}_{\sun}$ for [\OIII] and \Ha\ outflows respectively.

Assuming a biconical structure for the outflow \citep[e.g.,][]{2012MNRAS.425L..66M}, we estimate its mass outflow rate, energy outflow rate, and momentum outflow rate based on the outflow mass $M_{\rm out}$, velocity, and radius as follows:

\begin{equation}
\dot{M}_{\rm out} = \frac{3M_{\rm out} \nu_{\rm out}}{R_{\rm out}}
\end{equation}
\begin{equation}
\dot{E}_{\rm out} = \frac{1}{2} \dot{M}_{\rm out} \nu_{\rm out}^2 = \frac{3 M_{\rm out} \nu_{\rm out}^3}{2 R_{\rm out}}
\end{equation}
\begin{equation}
\dot{P}_{\rm out} = \dot{M}_{\rm out}\nu_{\rm out} = \frac{3M_{\rm gas} \nu_{\rm out}^2}{R_{\rm out}}.
\end{equation}

Under the same electron density assumption
and outflow radius conditions discussed previously, the derived properties of [\OIII] and \Ha\ outflows are listed in Table \ref{tab:table2}. In comparison with the AGN bolometric luminosity obtained from the SED analysis, we find that the energy coupling efficiency of the [\OIII] line is $\lesssim$ 1\%, similar to the results presented by \cite{2020ApJ...888..110J}. The observed outflow momentum substantially exceeds the expected momentum deposition rate \cite[$\sim1 \times 10^{36}$ dyne,][]{2021ApJ...919..122V} derived from the star formation rate of 130 $\mathrm{M}_{\sun}$/yr, providing  evidence for AGN-driven outflow. We note that for the  \Ha\ line, although its energy coupling efficiency is similar to that of  [\OIII], the measurement of this value is associated with significant uncertainties.

Assuming the outflow radius of 1.2 kpc as we discussed previously, the outflow timescale $t_{\rm outflow}$ needed for the outflow gas launched from the center of the host galaxy to its current observed position is 0.4-0.5 Myr for both [\OIII] and \Ha\ emitting gas. This timescale is a few times higher than the timescale based on the observed outflow mass to reach its mass outflow rate. However, we note that the outflow masses derived from the observed line luminosities are not corrected with the possible extinction, especially for the undetected outflow on the red and far side. If we take the extinction into account, the outflow timescales based on the physical extension and the mass are both $\sim$ a few $\times 10^{5}$ yr. This is similar to the outflow timescale estimated for luminous obscured AGN in \citet{2021ApJ...906...21J}, but it is shorter than the estimated timescale from radiation pressure. This suggests that radiation-driven outflows may be a possibility for multiple episodes of activity. For a more detailed discussion, see Section \ref{subsec:Driving Mechanisms}.

\subsection{BPT Diagram} \label{subsec:BPT Diagram}
 
\begin{figure}[!htbp]
\centering
\includegraphics[width=\columnwidth]{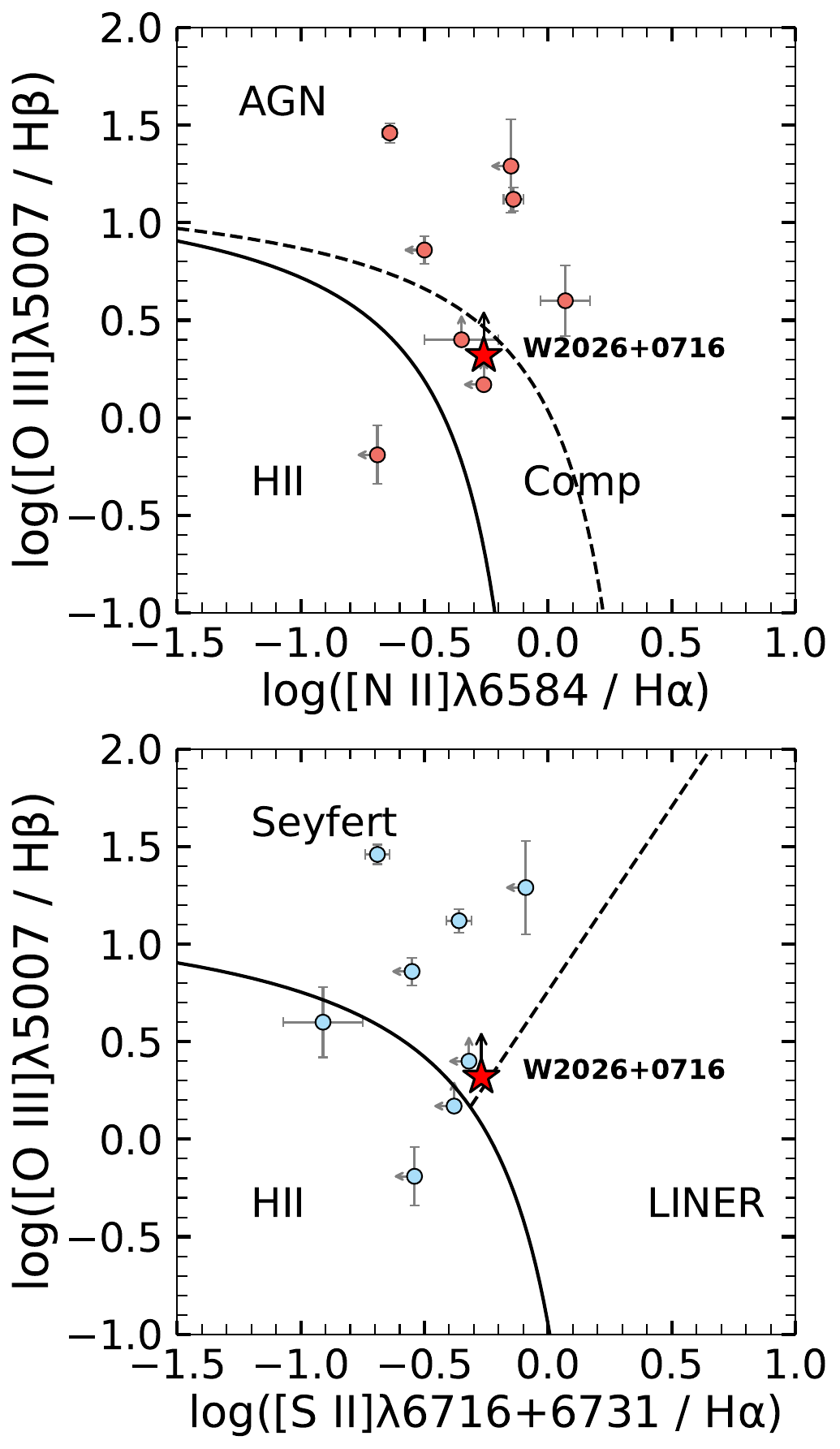}
\caption{\textbf{BPT diagram for \w2026.
} Colored dots represent the positions of Hot DOGs with known ionized gas outflows from the 1-D spectroscopic analysis\citep{2020ApJ...888..110J}, while the star symbol marks the position of W2026+0716, the subject of this study. Due to the lack of narrow 
 \Hb\
 line data, the broad-line 
 \Hb\
 flux is used instead, thus representing the lower limit of W2026+0716's position on the BPT diagram.}
\label{fig:BPT Diagram}
\end{figure}

Based on the narrow components in the spatially integrated spectrum, out
target source is located in the “composite”
region \citep{2003MNRAS.341...33K, 2013ApJ...774L..10K}
of the [\NII]/\Ha-vs-[\OIII]/\Hb\ version (see Figure \ref{fig:BPT Diagram}) of the BPT diagram \citep{1981PASP...93....5B}, suggesting the possible coexistence of AGN activity and star formation. In the [\SII]/\Ha-vs-[\OIII]/\Hb\ plot, \w2026 is positioned near the boundary between the Seyfert and LINER regions \citep{2006MNRAS.372..961K}, further supporting the interpretation of composite nature and indicating the presence of both high- and low-ionization gas. However, since only the broad-line component of \Hb\ could be measured, the arrow in the diagram represents a lower limit for the [\OIII]/\Hb\ ratio. The actual ratio could be higher, pushing this system further into the AGN region of the BPT diagram.

Compared to other Hot DOGs which range
from the AGN to composite regions on the BPT diagram, \w2026 exhibits typical characteristics, which indicates that the Hot DOGs generally have strong AGN activity, possibly accompanied by significant star formation \citep{2015ApJ...804...27A, 2021A&A...654A..37D}. 

\section{Discussion}\label{Discussion}
\subsection{Blue Hot DOGs} \label{subsec:Blue Hot DOGs}

\w2026\ exhibits excess AGN radiation in the optical blue and near-ultraviolet regions. The low-extinction AGN component contributes over 50\% of the blue luminosity below 1 \micron, clearly indicating \w2026\ as a Blue Hot DOG. This subpopulation, which may account for 26\% of all Hot DOGs \citep{2024ApJ...971...40L}, was first identified by \citet{2016ApJ...819..111A, 2020ApJ...897..112A}.

Polarimetric observations by \citet{2022ApJ...934..101A}
suggest that the blue light in the BHDs originates from the scattering of the central AGN's light. Among the first three Blue Hot DOGs identified, W0116$-$0505 and W0220$+$0137 have large-scale ionized gas outflows \citep{2020ApJ...905...16F}. In addition, W0116$-$0505 also exhibits extent broad CO emission lines \citep{2024MNRAS.534..978M}, suggesting an optically thick outflow scattering model. The spatially resolved emission lines found in this study of \w2026\  also reveal a strong ionized gas outflow. The complex outflow structure (as discussed in Section \ref{subsec:Kinematics Map}) might be an important feature of the optically thick outflow scattering model for BHDs.

\subsection{Multi Dynamical Components in the Biconical Outflows} \label{subsec:Biconical Outflows}\label{subsec:Multiphase Outflows}

As discussed in the Section \ref{subsec:Kinematics Map},
the broad components of [\OIII] and \Ha\
emission lines exhibit distinct features in their velocity dispersion maps. Both lines are spatially resolved in different directions in their flux intensity distribution, line-of-sight velocity shifts, and velocity dispersion.
High velocity dispersion regions are offset from the peak of the continuum emission, likely resulting from turbulent motions or shocks caused by the interaction between high-velocity outflows and the interstellar medium. These  results suggest that the spatial structure of the outflow is not uniformly spherical, but rather possesses a more complex geometric morphology.

\begin{figure}[t]
\centering
\includegraphics[width=\columnwidth]{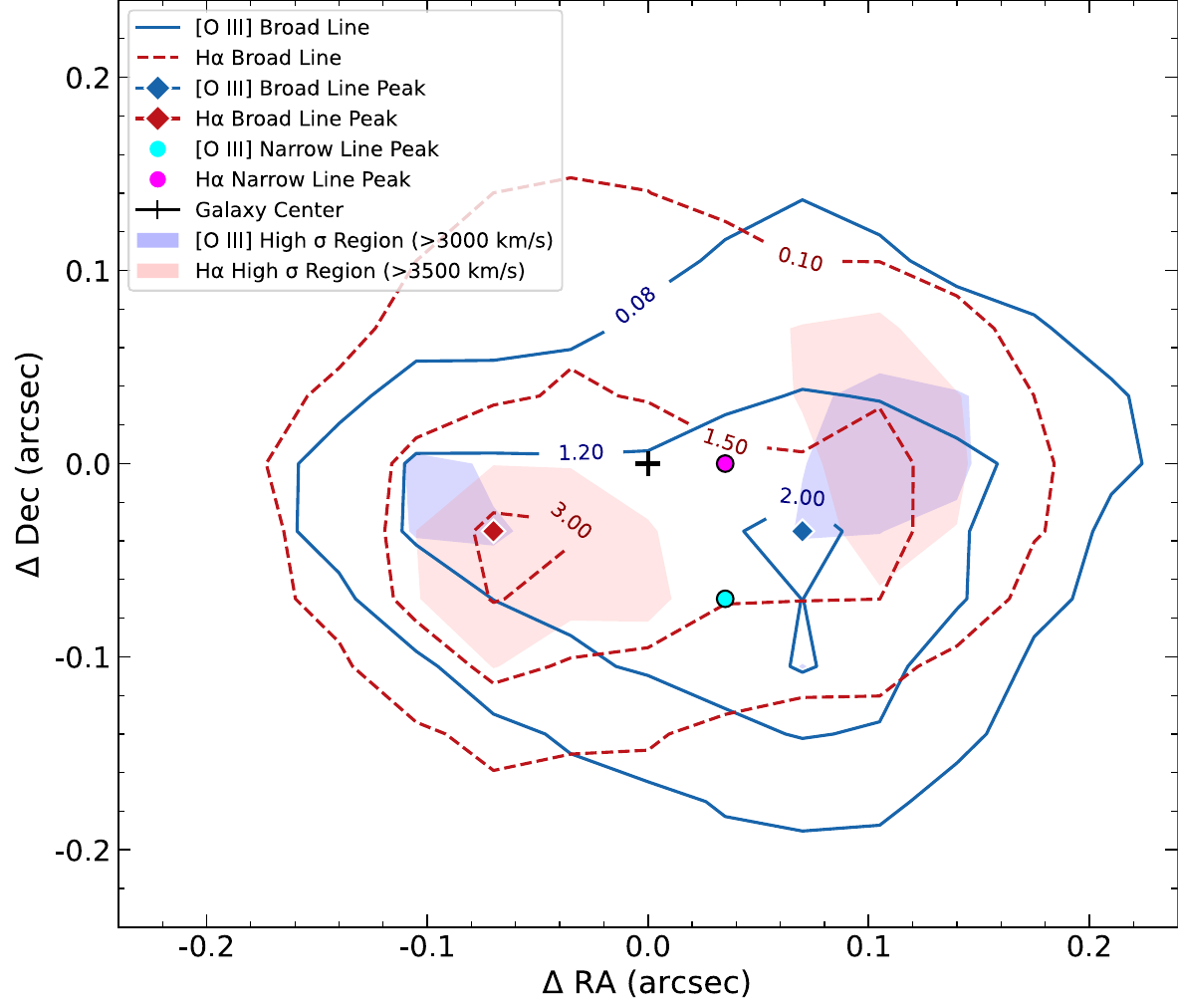}
\caption{
\textbf{Geometric distributions of the [\OIII], \Ha, and continuum emission.}
The blue contours represent the [\OIII] emission, while the red contours show the \Ha\ emission. The contours are labeled with the flux density in $10^{18}$ erg s$^{-1}$ cm$^{-2}$ \AA$^{-1}$. The peaks of the [\OIII], \Ha, and NIR continuum are presented with symbols in blue, red and black, respectively.
}
\label{fig:contour}
\end{figure}

\begin{figure*}[!htbp]
\centering
\includegraphics[width=\textwidth]{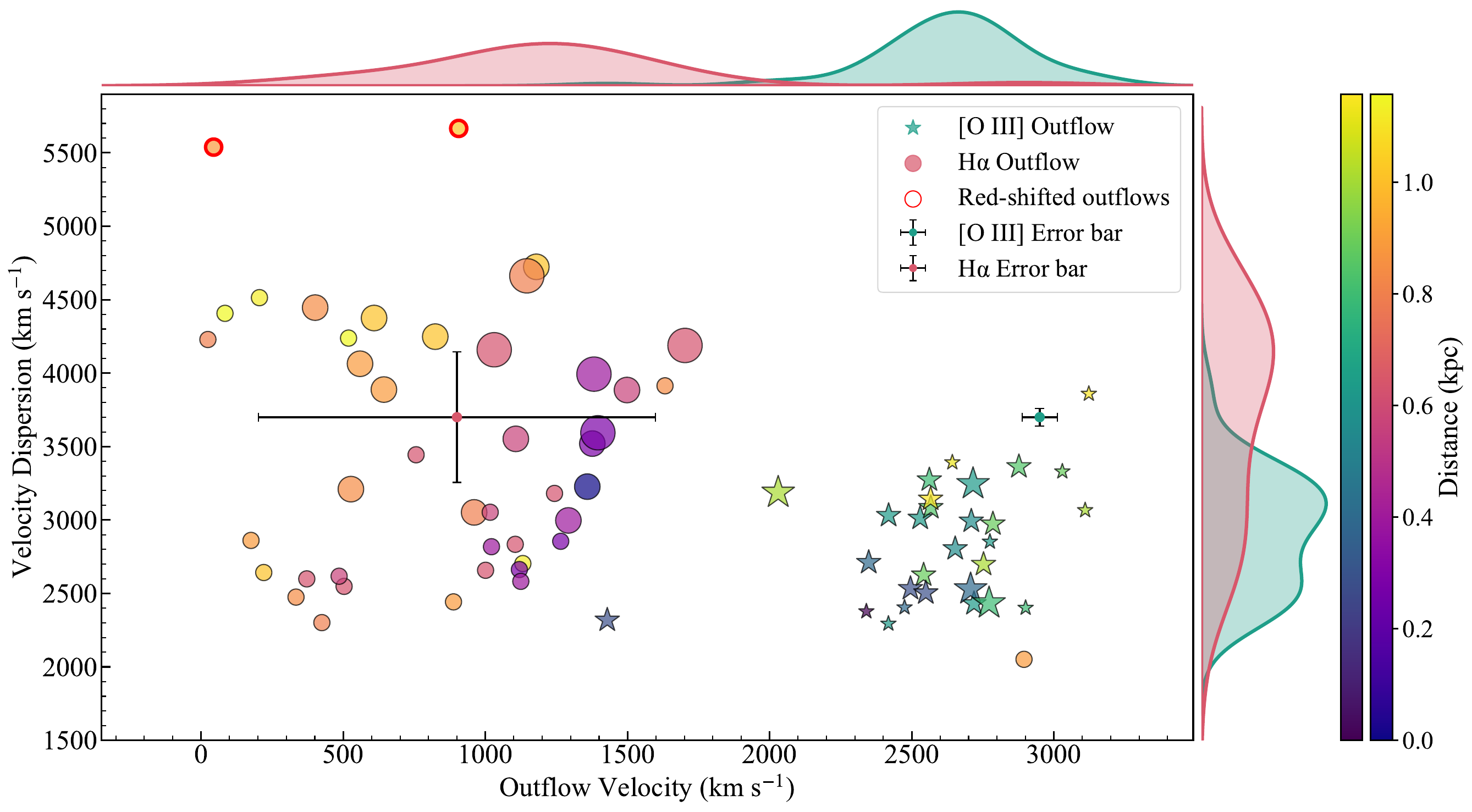}
\caption{\textbf{Velocity-velocity dispersion diagram of W2026+0716.} Circles and stars represent the outflow features of 
\Ha\
 and 
[\OIII], respectively, for each pixel. The color of each data point indicates the distance from the galactic center, while the size represents the relative flux.}
\label{fig:VVD}
\end{figure*}

Bipolar outflow models are often adopted to explain large-scale AGN outflow structures \citep{2005ARA&A..43..769V,husemann_reality_2018}. For the case of \w2026, we employ an inner-hollow bi-conical outflow model similar to that proposed by \citet{2016ApJ...828...97B}. In our system, the broad \Ha\ component exhibits an overall blueshift across the entire field of view, with significant line widths and non-uniform spatial flux distribution. These features strongly suggest that the broad \Ha\
component primarily originates from outflows. We consider the direction between the continuum peak to the center of the outflow flux as the primary outflow direction.

For both [\OIII] and \Ha\ line emission, the observed overall blueshifted velocity field indicates that the outflow is predominantly oriented towards the observer. Because Hot DOGs are a class of highly dust-obscured AGN galaxies, the surrounding dust significantly affects our observational results. Specifically, the redshifted side of the outflow has a longer path length along the line of sight and is thus more heavily obscured by dust. This effect makes it challenging to observe the redshifted portion of the outflows.

In the kinematics map analysis (Section \ref{subsec:Kinematics Map}), we found the spatial deviation between the outflow centers of [\OIII] and \Ha. Figure \ref{fig:contour} details the spatial distribution of [\OIII] and \Ha. Assuming the AGN is centrally located at the peak of continuum emission, we observe that the \Ha\ narrow line roughly coincides with the continuum peak within half a size of resolution element. However, the narrow line [\OIII] emission deviates from the AGN location by more than half a size of resolution element. This discrepancy may result from non-uniform dust extinction.

As shown in the velocity-velocity dispersion diagram
(Figure \ref{fig:VVD}), the [\OIII] outflow exhibits higher blueshifted velocities close to the center,
while \Ha\
shows relatively high velocity dispersion at larger radii. The inconsistency between \Ha\ and [\OIII]
outflow characteristics suggests a complex outflow structure in \w2026. Combined with the previously observed flux center asymmetry, this phenomenon leads us to speculate the [\OIII] and \Ha\
outflows may trace different components of the ionized gas outflow. [\OIII]
 and \Ha\
 have different critical densities, with \Ha\
more likely to exist in relatively cool, dense gas. Based on these observations, we hypothesize that the outflow may have a large opening angle with non-uniform gas density distribution and two primary outflow directions. The primary direction, traced primarily by [\OIII], 
represents the highest velocity component where gas density is relatively low and temperature is higher, favoring [\OIII] excitation. To verify this hypothesis, higher resolution observations are needed to reveal the spatial distribution of electron density.

\begin{figure}[t]
\centering
\includegraphics[width=\columnwidth]{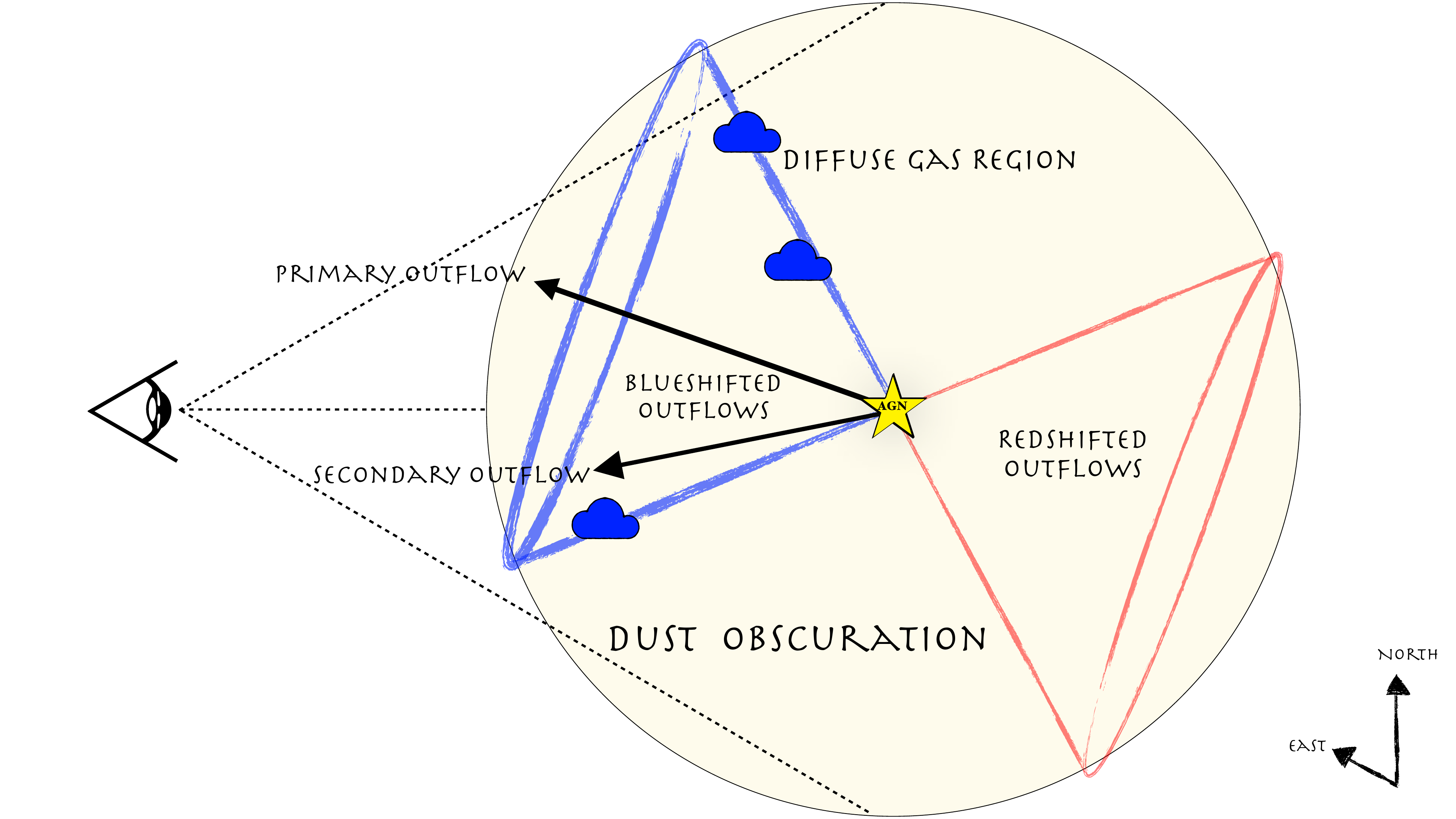}
\caption{\textbf{Schematic diagram of the bicone outflow model.} The model features inner and outer opening angles, with an assumed inclined bicone axis at $30\degree$
to the line of sight. The blue clouds in the outer opening angle region represent the possible presence of diffuse gas. The redshifted outflow is attenuated along the line of sight due to dust obscuration.}
\label{fig:bicone outflow model}
\end{figure}

As illustrated in Figure \ref{fig:bicone outflow model}, our proposed outflow model comprises two main components. The inner arrows represent the core of the outflow, demonstrating its primary direction and concentrated flux stream. Based on our assumption, the angle between the core outflow direction and the line of sight is $30^\circ$. The outer cone region represents slower velocity outflows or diffuse ionized gas.

\subsection{Role of Outflow in Obscured AGN Evolution}\label{subsec:Driving Mechanisms}

The outflows of two ionized gases display comparable dynamical characteristics. Both [\OIII] and \Ha\ outflows carry the kinetic energy outflow rates $\dot E_{\rm out} \sim M_{\rm out}v_{\rm out}^{3}/2R_{\rm out}$ of 1.6--1.7 $\times 10^{45}$\,erg\,s$^{-1}$ and momentum outflow rates $\dot P_{\rm out} \sim M_{\rm out}v_{\rm out}^{2}/R_{\rm out}$ of 1.0--1.4$\times 10^{37}$\, dyn. With their similar inferred energy coupling efficiencies $\dot E_{\rm out}/L_{\rm AGN} \sim 0.6\%$ and  momentum boost factor \citep{2014MNRAS.444.2355C} $\dot P_{\rm out}\,c/L_{\rm AGN} \gtrsim 1$, these two ionized gas outflows yet exhibit noticeable differences in the spatial distributions of their morphology and kinematics. This discrepancy may indicate that the [\OIII] and \Ha\ emission lines trace distinct components of ionized outflows driven by a common physical mechanism, rather than by separate processes \citep[e.g., AGN and starburst activities in][]{2022ApJ...933..110X}. This suggests that multiple kinematically distinct ionized outflow components may coexist in this system. Previous studies have primarily focused on multiphase outflows involving ionized, neutral, or molecular gas \citep[e.g.,][]{2021MNRAS.505.5753F}, whereas this work provides the potential first evidence of kinematic differentiation within the ionized outflow itself. The observed differentiation might result from varying electron densities in high- and low-speed outflows. This finding suggests that the outflow structure at this evolutionary stage may be more complex than previously thought. Such a phenomenon could be associated with outflow fragmentation or differential acceleration, where ionized gas components with different physical properties have not yet fully mixed. However, given the limitations of our data in terms of signal-to-noise ratio, as well as potential contamination from BLR scattered light and dust extinction, our results provide only preliminary evidence for this scenario. Further observations with higher spatial and spectral resolution are required to confirm this phenomenon and to gain a deeper understanding of its implications for AGN feedback.

\citet{2024ApJ...971...40L} based on number density comparisons and Type 1 AGN lifetimes, estimate the lifespan of Blue Hot DOGs to be approximately 0.5 Myr and that of ordinary Hot DOGs to be about 1.5 Myr. These timescales are comparable to the calculated outflow timescale $t_{\rm outflow}$(see discussions in Section \ref{subsec:Outflows Energetics Calculation}). This
implies that outflows likely persist throughout the Hot DOGs phase, which is potential evidence for the widespread presence of outflows during Hot DOGs stage.
The kinematic timescales of the [\OIII] and \Ha\
outflows are similar, suggesting that they may originate from the same epoch. Therefore, detecting multi-component outflows and assessing the long-term evolution of outflow kinematics will provide crucial information about the transitional phase between obscured and unobscured AGN systems.

\section{ Conclusion}\label{sec:Conclusion}

We conducted narrow \textit{$Hn5$} band and \textit{$Kn5$} band IFU observations of \w2026
using Keck/OSIRIS. The observational results reveal extremely broad [\OIII] and \Ha\
emission lines. 
We provided detailed kinematic and dynamical analyses of the broad and narrow components of these lines.
Additionally, we performed multi-parameter SED
fitting for W2026$+$0716. Our analyses have found:

\begin{enumerate}
    \item \w2026\ is identified as a Blue Hot DOG exhibiting strong outflows. Blue Hot DOGs may universally host outflows, which could play a crucial role in cleaning out the dust cocoon around the central accreting supermassive black hole to transit to an unobscured AGN.

    \item Kinematic analysis of \w2026\ suggests that the bulk
outflow has a small inclination angle to the line of sight. The geometric distribution of outflow regions with the high velocity and high dispersion suggest that this outflow is likely a bi-conical structure with some complexity. The extinction of dust surrounding the central region of this Hot DOG hinders the detections of
the redshifted outflow on the far side.

    \item The [\OIII] and \Ha\
outflows exhibit different spatial distributions and velocity dispersion profiles, likely tracing gas at different densities and ionization states. Consequently, W2026$+$0716 may possess a complex multi-component ionized gas outflow structure. The broad \Ha\ line from the BLR is not clearly detected, possibly affected by the high dust obscuration toward the central AGN.

    \item Based on the timescale estimates,
both ionized gas outflows have persisted over a few $\times 10^{5}$ yr within the evolutionary phase of the Hot DOG W2026$+$0716. This duration is comparable to the lifetime of Blue Hot DOGs. 
\end{enumerate}

The outflow kinematics and geometry of the obscured AGN provide critical clues about the process of how the dust cocoon is revealed. Furthermore, the relative configuration of the outflows and the ISM distribution in the galaxy could also lead to positive or negative AGN feedback to the star formation activities in the host system. The spatially resolved spectroscopic studies for the obscured AGN systems, based on the IFU instrumentations (Keck, VLT, or \textit{JWST} for intermediate redshift targets, while \textit{CSST}\citep{2011SSPMA..41.1441Z} for low redshift sources), can better provide a comprehensive picture of the critical transitional phase between the obscured and unobscured AGNs.

\section*{Acknowledgments}\label{sec:acknowledgments}
This material is based upon work supported by the National Natural Science Foundation of China (grant No. 11988101, 11973051). C.-W.T. and M.L. are supported by the International Partnership Program of the Chinese Academy of Sciences, Grant No. 114A11KYSB20210010. RJA was supported by FONDECYT grant number 1231718 and by the ANID BASAL project FB210003.
This publication makes use of data obtained at the W.M. Keck Observatory, which is operated as a scientific partnership among Caltech, the University of California and NASA. The Keck Observatory was made possible by the generous financial support of the W.M. Keck Foundation. 
The authors wish to recognize and acknowledge the very significant cultural role and reverence that the summit of Mauna Kea has always had within the indigenous Hawaiian community. We are most fortunate to have the opportunity to conduct observations from this mountain.
This publication makes use of data products from the {\it Wide-field Infrared Survey Explorer}, which is a joint project of the University of California, Los Angeles, and the Jet Propulsion Laboratory, California Institute of Technology, funded by the National Aeronautics and Space Administration. 
This research has made use of the NASA/ IPAC Infrared Science Archive, which is operated by the Jet Propulsion Laboratory, California Institute of Technology, under ontract with the National Aeronautics and Space Administration. All \textit{HST}, Pan-STARRS1, and \textit{Spitzer} data referenced in this paper are accessible through the Mikulski Archive for Space Telescopes (MAST) at \href{https://archive.stsci.edu/doi/resolve/resolve.html?doi=10.17909/y5pw-cx63}{DOI: 10.17909/y5pw-cx63}, while AllWISE data can be accessed through IPAC via \href{https://catcopy.ipac.caltech.edu/dois/doi.php?id=10.26131/IRSA1}{DOI: 10.26131/IRSA1}.

\facility{Keck:I (OSIRIS); \textit{Spitzer Space Telescope}; \textit{Herschel Space Telescope}; \textit{Hubble Space Telescope}; \textit{Wide-field Infrared Survey Explorer}; JCMT} \\

\bibliography{ref.bib}

\begin{thebibliography}{}
\expandafter\ifx\csname natexlab\endcsname\relax\def\natexlab#1{#1}\fi
\providecommand{\url}[1]{\href{#1}{#1}}
\providecommand{\dodoi}[1]{doi:~\href{http://doi.org/#1}{\nolinkurl{#1}}}
\providecommand{\doeprint}[1]{\href{http://ascl.net/#1}{\nolinkurl{http://ascl.net/#1}}}
\providecommand{\doarXiv}[1]{\href{https://arxiv.org/abs/#1}{\nolinkurl{https://arxiv.org/abs/#1}}}

\bibitem[{{Assef} {et~al.}(2010){Assef}, {Kochanek}, {Brodwin}, {Cool}, {Forman}, {Gonzalez}, {Hickox}, {Jones}, {Le Floc'h}, {Moustakas}, {Murray}, \& {Stern}}]{2010ApJ...713..970A}
{Assef}, R.~J., {Kochanek}, C.~S., {Brodwin}, M., {et~al.} 2010, \apj, 713, 970, \dodoi{10.1088/0004-637X/713/2/970}

\bibitem[{{Assef} {et~al.}(2015){Assef}, {Eisenhardt}, {Stern}, {Tsai}, {Wu}, {Wylezalek}, {Blain}, {Bridge}, {Donoso}, {Gonzales}, {Griffith}, \& {Jarrett}}]{2015ApJ...804...27A}
{Assef}, R.~J., {Eisenhardt}, P.~R.~M., {Stern}, D., {et~al.} 2015, \apj, 804, 27, \dodoi{10.1088/0004-637X/804/1/27}

\bibitem[{{Assef} {et~al.}(2016){Assef}, {Walton}, {Brightman}, {Stern}, {Alexander}, {Bauer}, {Blain}, {Diaz-Santos}, {Eisenhardt}, {Finkelstein}, {Hickox}, {Tsai}, \& {Wu}}]{2016ApJ...819..111A}
{Assef}, R.~J., {Walton}, D.~J., {Brightman}, M., {et~al.} 2016, \apj, 819, 111, \dodoi{10.3847/0004-637X/819/2/111}

\bibitem[{{Assef} {et~al.}(2020){Assef}, {Brightman}, {Walton}, {Stern}, {Bauer}, {Blain}, {D{\'\i}az-Santos}, {Eisenhardt}, {Hickox}, {Jun}, {Psychogyios}, {Tsai}, \& {Wu}}]{2020ApJ...897..112A}
{Assef}, R.~J., {Brightman}, M., {Walton}, D.~J., {et~al.} 2020, \apj, 897, 112, \dodoi{10.3847/1538-4357/ab9814}

\bibitem[{{Assef} {et~al.}(2022){Assef}, {Bauer}, {Blain}, {Brightman}, {D{\'\i}az-Santos}, {Eisenhardt}, {Jun}, {Stern}, {Tsai}, {Walton}, \& {Wu}}]{2022ApJ...934..101A}
{Assef}, R.~J., {Bauer}, F.~E., {Blain}, A.~W., {et~al.} 2022, \apj, 934, 101, \dodoi{10.3847/1538-4357/ac77fc}

\bibitem[{{Bae} \& {Woo}(2016)}]{2016ApJ...828...97B}
{Bae}, H.-J., \& {Woo}, J.-H. 2016, \apj, 828, 97, \dodoi{10.3847/0004-637X/828/2/97}

\bibitem[{{Baldwin} {et~al.}(1981){Baldwin}, {Phillips}, \& {Terlevich}}]{1981PASP...93....5B}
{Baldwin}, J.~A., {Phillips}, M.~M., \& {Terlevich}, R. 1981, \pasp, 93, 5, \dodoi{10.1086/130766}

\bibitem[{{Blain}(2003)}]{2003ASPC..289..247B}
{Blain}, A.~W. 2003, in Astronomical Society of the Pacific Conference Series, Vol. 289, The Proceedings of the IAU 8th Asian-Pacific Regional Meeting, Volume 1, ed. S.~{Ikeuchi}, J.~{Hearnshaw}, \& T.~{Hanawa}, 247--250, \dodoi{10.48550/arXiv.astro-ph/0302184}

\bibitem[{{Bussmann} {et~al.}(2009){Bussmann}, {Dey}, {Lotz}, {Armus}, {Brand}, {Brown}, {Desai}, {Eisenhardt}, {Higdon}, {Higdon}, {Jannuzi}, {Le Floc'h}, {Melbourne}, {Soifer}, \& {Weedman}}]{2009ApJ...693..750B}
{Bussmann}, R.~S., {Dey}, A., {Lotz}, J., {et~al.} 2009, \apj, 693, 750, \dodoi{10.1088/0004-637X/693/1/750}

\bibitem[{{Calistro Rivera} {et~al.}(2021){Calistro Rivera}, {Alexander}, {Rosario}, {Harrison}, {Stalevski}, {Rakshit}, {Fawcett}, {Morabito}, {Klindt}, {Best}, {Bonato}, {Bowler}, {Costa}, \& {Kondapally}}]{2021A&A...649A.102C}
{Calistro Rivera}, G., {Alexander}, D.~M., {Rosario}, D.~J., {et~al.} 2021, \aap, 649, A102, \dodoi{10.1051/0004-6361/202040214}

\bibitem[{{Canalizo} \& {Stockton}(2001)}]{2001ApJ...555..719C}
{Canalizo}, G., \& {Stockton}, A. 2001, \apj, 555, 719, \dodoi{10.1086/321520}

\bibitem[{{Carniani} {et~al.}(2015){Carniani}, {Marconi}, {Maiolino}, {Balmaverde}, {Brusa}, {Cano-D{\'\i}az}, {Cicone}, {Comastri}, {Cresci}, {Fiore}, {Feruglio}, {La Franca}, {Mainieri}, {Mannucci}, {Nagao}, {Netzer}, {Piconcelli}, {Risaliti}, {Schneider}, \& {Shemmer}}]{2015A&A...580A.102C}
{Carniani}, S., {Marconi}, A., {Maiolino}, R., {et~al.} 2015, \aap, 580, A102, \dodoi{10.1051/0004-6361/201526557}

\bibitem[{{Casey} {et~al.}(2014){Casey}, {Narayanan}, \& {Cooray}}]{2014PhR...541...45C}
{Casey}, C.~M., {Narayanan}, D., \& {Cooray}, A. 2014, \physrep, 541, 45, \dodoi{10.1016/j.physrep.2014.02.009}

\bibitem[{{Chambers} {et~al.}(2016){Chambers}, {Magnier}, {Metcalfe}, {Flewelling}, {Huber}, {Waters}, {Denneau}, {Draper}, {Farrow}, {Finkbeiner}, {Holmberg}, {Koppenhoefer}, {Price}, {Rest}, {Saglia}, {Schlafly}, {Smartt}, {Sweeney}, {Wainscoat}, {Burgett}, {Chastel}, {Grav}, {Heasley}, {Hodapp}, {Jedicke}, {Kaiser}, {Kudritzki}, {Luppino}, {Lupton}, {Monet}, {Morgan}, {Onaka}, {Shiao}, {Stubbs}, {Tonry}, {White}, {Ba{\~n}ados}, {Bell}, {Bender}, {Bernard}, {Boegner}, {Boffi}, {Botticella}, {Calamida}, {Casertano}, {Chen}, {Chen}, {Cole}, {Deacon}, {Frenk}, {Fitzsimmons}, {Gezari}, {Gibbs}, {Goessl}, {Goggia}, {Gourgue}, {Goldman}, {Grant}, {Grebel}, {Hambly}, {Hasinger}, {Heavens}, {Heckman}, {Henderson}, {Henning}, {Holman}, {Hopp}, {Ip}, {Isani}, {Jackson}, {Keyes}, {Koekemoer}, {Kotak}, {Le}, {Liska}, {Long}, {Lucey}, {Liu}, {Martin}, {Masci}, {McLean}, {Mindel}, {Misra}, {Morganson}, {Murphy}, {Obaika}, {Narayan}, {Nieto-Santisteban}, {Norberg}, {Peacock}, {Pier}, {Postman}, {Primak}, {Rae}, {Rai},
  {Riess}, {Riffeser}, {Rix}, {R{\"o}ser}, {Russel}, {Rutz}, {Schilbach}, {Schultz}, {Scolnic}, {Strolger}, {Szalay}, {Seitz}, {Small}, {Smith}, {Soderblom}, {Taylor}, {Thomson}, {Taylor}, {Thakar}, {Thiel}, {Thilker}, {Unger}, {Urata}, {Valenti}, {Wagner}, {Walder}, {Walter}, {Watters}, {Werner}, {Wood-Vasey}, \& {Wyse}}]{2016arXiv161205560C}
{Chambers}, K.~C., {Magnier}, E.~A., {Metcalfe}, N., {et~al.} 2016, arXiv e-prints, arXiv:1612.05560, \dodoi{10.48550/arXiv.1612.05560}

\bibitem[{{Cicone} {et~al.}(2018){Cicone}, {Brusa}, {Ramos Almeida}, {Cresci}, {Husemann}, \& {Mainieri}}]{2018NatAs...2..176C}
{Cicone}, C., {Brusa}, M., {Ramos Almeida}, C., {et~al.} 2018, Nature Astronomy, 2, 176, \dodoi{10.1038/s41550-018-0406-3}

\bibitem[{{Cicone} {et~al.}(2014){Cicone}, {Maiolino}, {Sturm}, {Graci{\'a}-Carpio}, {Feruglio}, {Neri}, {Aalto}, {Davies}, {Fiore}, {Fischer}, {Garc{\'\i}a-Burillo}, {Gonz{\'a}lez-Alfonso}, {Hailey-Dunsheath}, {Piconcelli}, \& {Veilleux}}]{2014A&A...562A..21C}
{Cicone}, C., {Maiolino}, R., {Sturm}, E., {et~al.} 2014, \aap, 562, A21, \dodoi{10.1051/0004-6361/201322464}

\bibitem[{{Costa} {et~al.}(2014){Costa}, {Sijacki}, \& {Haehnelt}}]{2014MNRAS.444.2355C}
{Costa}, T., {Sijacki}, D., \& {Haehnelt}, M.~G. 2014, \mnras, 444, 2355, \dodoi{10.1093/mnras/stu1632}

\bibitem[{{Cutri} {et~al.}(2021){Cutri}, {Wright}, {Conrow}, {Fowler}, {Eisenhardt}, {Grillmair}, {Kirkpatrick}, {Masci}, {McCallon}, {Wheelock}, {Fajardo-Acosta}, {Yan}, {Benford}, {Harbut}, {Jarrett}, {Lake}, {Leisawitz}, {Ressler}, {Stanford}, {Tsai}, {Liu}, {Helou}, {Mainzer}, {Gettngs}, {Gonzalez}, {Hoffman}, {Marsh}, {Padgett}, {Skrutskie}, {Beck}, {Papin}, \& {Wittman}}]{2014yCat.2328....0C}
{Cutri}, R.~M., {Wright}, E.~L., {Conrow}, T., {et~al.} 2021, {VizieR Online Data Catalog: AllWISE Data Release (Cutri+ 2013)}, VizieR On-line Data Catalog: II/328. Originally published in: IPAC/Caltech (2013)

\bibitem[{{Davies}(2007)}]{2007MNRAS.375.1099D}
{Davies}, R.~I. 2007, \mnras, 375, 1099, \dodoi{10.1111/j.1365-2966.2006.11383.x}

\bibitem[{{Di Matteo} {et~al.}(2005){Di Matteo}, {Springel}, \& {Hernquist}}]{2005Natur.433..604D}
{Di Matteo}, T., {Springel}, V., \& {Hernquist}, L. 2005, \nat, 433, 604, \dodoi{10.1038/nature03335}

\bibitem[{{D{\'\i}az-Santos} {et~al.}(2018){D{\'\i}az-Santos}, {Assef}, {Blain}, {Aravena}, {Stern}, {Tsai}, {Eisenhardt}, {Wu}, {Jun}, {Dibert}, {Inami}, {Lansbury}, \& {Leclercq}}]{2018Sci...362.1034D}
{D{\'\i}az-Santos}, T., {Assef}, R.~J., {Blain}, A.~W., {et~al.} 2018, Science, 362, 1034, \dodoi{10.1126/science.aap7605}

\bibitem[{{D{\'\i}az-Santos} {et~al.}(2021){D{\'\i}az-Santos}, {Assef}, {Eisenhardt}, {Jun}, {Jones}, {Blain}, {Stern}, {Aravena}, {Tsai}, {Lake}, {Wu}, \& {Gonz{\'a}lez-L{\'o}pez}}]{2021A&A...654A..37D}
{D{\'\i}az-Santos}, T., {Assef}, R.~J., {Eisenhardt}, P. R.~M., {et~al.} 2021, \aap, 654, A37, \dodoi{10.1051/0004-6361/202140455}

\bibitem[{{Eisenhardt} {et~al.}(2012){Eisenhardt}, {Wu}, {Tsai}, {Assef}, {Benford}, {Blain}, {Bridge}, {Condon}, {Cushing}, {Cutri}, {Evans}, {Gelino}, {Griffith}, {Grillmair}, {Jarrett}, {Lonsdale}, {Masci}, {Mason}, {Petty}, {Sayers}, {Stanford}, {Stern}, {Wright}, \& {Yan}}]{2012ApJ...755..173E}
{Eisenhardt}, P. R.~M., {Wu}, J., {Tsai}, C.-W., {et~al.} 2012, \apj, 755, 173, \dodoi{10.1088/0004-637X/755/2/173}

\bibitem[{{Fabian}(2012)}]{2012ARA&A..50..455F}
{Fabian}, A.~C. 2012, \araa, 50, 455, \dodoi{10.1146/annurev-astro-081811-125521}

\bibitem[{{Fan} {et~al.}(2016){Fan}, {Han}, {Fang}, {Gao}, {Zhang}, {Jiang}, {Wu}, {Yang}, \& {Li}}]{2016ApJ...822L..32F}
{Fan}, L., {Han}, Y., {Fang}, G., {et~al.} 2016, \apjl, 822, L32, \dodoi{10.3847/2041-8205/822/2/L32}

\bibitem[{{Faucher-Gigu{\`e}re} \& {Quataert}(2012)}]{2012MNRAS.425..605F}
{Faucher-Gigu{\`e}re}, C.-A., \& {Quataert}, E. 2012, \mnras, 425, 605, \dodoi{10.1111/j.1365-2966.2012.21512.x}

\bibitem[{{Feruglio} {et~al.}(2010){Feruglio}, {Maiolino}, {Piconcelli}, {Menci}, {Aussel}, {Lamastra}, \& {Fiore}}]{2010A&A...518L.155F}
{Feruglio}, C., {Maiolino}, R., {Piconcelli}, E., {et~al.} 2010, \aap, 518, L155, \dodoi{10.1051/0004-6361/201015164}

\bibitem[{{Feruglio} {et~al.}(2015){Feruglio}, {Fiore}, {Carniani}, {Piconcelli}, {Zappacosta}, {Bongiorno}, {Cicone}, {Maiolino}, {Marconi}, {Menci}, {Puccetti}, \& {Veilleux}}]{2015A&A...583A..99F}
{Feruglio}, C., {Fiore}, F., {Carniani}, S., {et~al.} 2015, \aap, 583, A99, \dodoi{10.1051/0004-6361/201526020}

\bibitem[{{Finnerty} {et~al.}(2020){Finnerty}, {Larson}, {Soifer}, {Armus}, {Matthews}, {Jun}, {Moon}, {Melbourne}, {Gomez}, {Tsai}, {D{\'\i}az-Santos}, {Eisenhardt}, \& {Cushing}}]{2020ApJ...905...16F}
{Finnerty}, L., {Larson}, K., {Soifer}, B.~T., {et~al.} 2020, \apj, 905, 16, \dodoi{10.3847/1538-4357/abc3bf}

\bibitem[{{Fiore} {et~al.}(2017){Fiore}, {Feruglio}, {Shankar}, {Bischetti}, {Bongiorno}, {Brusa}, {Carniani}, {Cicone}, {Duras}, {Lamastra}, {Mainieri}, {Marconi}, {Menci}, {Maiolino}, {Piconcelli}, {Vietri}, \& {Zappacosta}}]{2017A&A...601A.143F}
{Fiore}, F., {Feruglio}, C., {Shankar}, F., {et~al.} 2017, \aap, 601, A143, \dodoi{10.1051/0004-6361/201629478}

\bibitem[{{Flewelling} {et~al.}(2020){Flewelling}, {Magnier}, {Chambers}, {Heasley}, {Holmberg}, {Huber}, {Sweeney}, {Waters}, {Calamida}, {Casertano}, {Chen}, {Farrow}, {Hasinger}, {Henderson}, {Long}, {Metcalfe}, {Narayan}, {Nieto-Santisteban}, {Norberg}, {Rest}, {Saglia}, {Szalay}, {Thakar}, {Tonry}, {Valenti}, {Werner}, {White}, {Denneau}, {Draper}, {Hodapp}, {Jedicke}, {Kaiser}, {Kudritzki}, {Price}, {Wainscoat}, {Chastel}, {McLean}, {Postman}, \& {Shiao}}]{2020ApJS..251....7F}
{Flewelling}, H.~A., {Magnier}, E.~A., {Chambers}, K.~C., {et~al.} 2020, \apjs, 251, 7, \dodoi{10.3847/1538-4365/abb82d}

\bibitem[{{Fluetsch} {et~al.}(2021){Fluetsch}, {Maiolino}, {Carniani}, {Arribas}, {Belfiore}, {Bellocchi}, {Cazzoli}, {Cicone}, {Cresci}, {Fabian}, {Gallagher}, {Ishibashi}, {Mannucci}, {Marconi}, {Perna}, {Sturm}, \& {Venturi}}]{2021MNRAS.505.5753F}
{Fluetsch}, A., {Maiolino}, R., {Carniani}, S., {et~al.} 2021, \mnras, 505, 5753, \dodoi{10.1093/mnras/stab1666}

\bibitem[{{Ginolfi} {et~al.}(2022){Ginolfi}, {Piconcelli}, {Zappacosta}, {Jones}, {Pentericci}, {Maiolino}, {Travascio}, {Menci}, {Carniani}, {Rizzo}, {Arrigoni Battaia}, {Cantalupo}, {De Breuck}, {Graziani}, {Knudsen}, {Laursen}, {Mainieri}, {Schneider}, {Stanley}, {Valiante}, \& {Verhamme}}]{2022NatCo..13.4574G}
{Ginolfi}, M., {Piconcelli}, E., {Zappacosta}, L., {et~al.} 2022, Nature Communications, 13, 4574, \dodoi{10.1038/s41467-022-32297-x}

\bibitem[{{Glikman} {et~al.}(2022){Glikman}, {Lacy}, {LaMassa}, {Bradley}, {Djorgovski}, {Urrutia}, {Gates}, {Graham}, {Urry}, \& {Yoon}}]{2022ApJ...934..119G}
{Glikman}, E., {Lacy}, M., {LaMassa}, S., {et~al.} 2022, \apj, 934, 119, \dodoi{10.3847/1538-4357/ac6bee}

\bibitem[{{Greene} {et~al.}(2011){Greene}, {Zakamska}, {Ho}, \& {Barth}}]{2011ApJ...732....9G}
{Greene}, J.~E., {Zakamska}, N.~L., {Ho}, L.~C., \& {Barth}, A.~J. 2011, \apj, 732, 9, \dodoi{10.1088/0004-637X/732/1/9}

\bibitem[{{Griffith} {et~al.}(2012){Griffith}, {Kirkpatrick}, {Eisenhardt}, {Gelino}, {Cushing}, {Benford}, {Blain}, {Bridge}, {Cohen}, {Cutri}, {Donoso}, {Jarrett}, {Lonsdale}, {Mace}, {Mainzer}, {Marsh}, {Padgett}, {Petty}, {Ressler}, {Skrutskie}, {Stanford}, {Stern}, {Tsai}, {Wright}, {Wu}, \& {Yan}}]{2012AJ....144..148G}
{Griffith}, R.~L., {Kirkpatrick}, J.~D., {Eisenhardt}, P. R.~M., {et~al.} 2012, \aj, 144, 148, \dodoi{10.1088/0004-6256/144/5/148}

\bibitem[{{Guo} {et~al.}(2018){Guo}, {Shen}, \& {Wang}}]{2018ascl.soft09008G}
{Guo}, H., {Shen}, Y., \& {Wang}, S. 2018, {PyQSOFit: Python code to fit the spectrum of quasars}, Astrophysics Source Code Library, record ascl:1809.008

\bibitem[{{Harrison} {et~al.}(2014){Harrison}, {Alexander}, {Mullaney}, \& {Swinbank}}]{2014MNRAS.441.3306H}
{Harrison}, C.~M., {Alexander}, D.~M., {Mullaney}, J.~R., \& {Swinbank}, A.~M. 2014, \mnras, 441, 3306, \dodoi{10.1093/mnras/stu515}

\bibitem[{{Heckman} \& {Best}(2014)}]{2014ARA&A..52..589H}
{Heckman}, T.~M., \& {Best}, P.~N. 2014, \araa, 52, 589, \dodoi{10.1146/annurev-astro-081913-035722}

\bibitem[{{Hopkins} {et~al.}(2016){Hopkins}, {Torrey}, {Faucher-Gigu{\`e}re}, {Quataert}, \& {Murray}}]{2016MNRAS.458..816H}
{Hopkins}, P.~F., {Torrey}, P., {Faucher-Gigu{\`e}re}, C.-A., {Quataert}, E., \& {Murray}, N. 2016, \mnras, 458, 816, \dodoi{10.1093/mnras/stw289}

\bibitem[{HSA(2021)}]{Herschel2021}
HSA. 2021, Herschel High Level Images,  IPAC, \dodoi{10.26131/IRSA79}

\bibitem[{Husemann \& Harrison(2018)}]{husemann_reality_2018}
Husemann, B., \& Harrison, C.~M. 2018, Nature Astronomy, 2, 196, \dodoi{10.1038/s41550-018-0407-2}

\bibitem[{{Jones} {et~al.}(2014){Jones}, {Blain}, {Stern}, {Assef}, {Bridge}, {Eisenhardt}, {Petty}, {Wu}, {Tsai}, {Cutri}, {Wright}, \& {Yan}}]{2014MNRAS.443..146J}
{Jones}, S.~F., {Blain}, A.~W., {Stern}, D., {et~al.} 2014, \mnras, 443, 146, \dodoi{10.1093/mnras/stu1157}

\bibitem[{{Jun} {et~al.}(2021){Jun}, {Assef}, {Carroll}, {Hickox}, {Kim}, {Lee}, {Ricci}, \& {Stern}}]{2021ApJ...906...21J}
{Jun}, H.~D., {Assef}, R.~J., {Carroll}, C.~M., {et~al.} 2021, \apj, 906, 21, \dodoi{10.3847/1538-4357/abc629}

\bibitem[{{Jun} {et~al.}(2020){Jun}, {Assef}, {Bauer}, {Blain}, {D{\'\i}az-Santos}, {Eisenhardt}, {Stern}, {Tsai}, {Wright}, \& {Wu}}]{2020ApJ...888..110J}
{Jun}, H.~D., {Assef}, R.~J., {Bauer}, F.~E., {et~al.} 2020, \apj, 888, 110, \dodoi{10.3847/1538-4357/ab5e7b}

\bibitem[{{Karouzos} {et~al.}(2016){Karouzos}, {Woo}, \& {Bae}}]{2016ApJ...833..171K}
{Karouzos}, M., {Woo}, J.-H., \& {Bae}, H.-J. 2016, \apj, 833, 171, \dodoi{10.3847/1538-4357/833/2/171}

\bibitem[{{Kauffmann} {et~al.}(2003){Kauffmann}, {Heckman}, {White}, {Charlot}, {Tremonti}, {Brinchmann}, {Bruzual}, {Peng}, {Seibert}, {Bernardi}, {Blanton}, {Brinkmann}, {Castander}, {Cs{\'a}bai}, {Fukugita}, {Ivezic}, {Munn}, {Nichol}, {Padmanabhan}, {Thakar}, {Weinberg}, \& {York}}]{2003MNRAS.341...33K}
{Kauffmann}, G., {Heckman}, T.~M., {White}, S. D.~M., {et~al.} 2003, \mnras, 341, 33, \dodoi{10.1046/j.1365-8711.2003.06291.x}

\bibitem[{{Kewley} {et~al.}(2006){Kewley}, {Groves}, {Kauffmann}, \& {Heckman}}]{2006MNRAS.372..961K}
{Kewley}, L.~J., {Groves}, B., {Kauffmann}, G., \& {Heckman}, T. 2006, \mnras, 372, 961, \dodoi{10.1111/j.1365-2966.2006.10859.x}

\bibitem[{{Kewley} {et~al.}(2013){Kewley}, {Maier}, {Yabe}, {Ohta}, {Akiyama}, {Dopita}, \& {Yuan}}]{2013ApJ...774L..10K}
{Kewley}, L.~J., {Maier}, C., {Yabe}, K., {et~al.} 2013, \apjl, 774, L10, \dodoi{10.1088/2041-8205/774/1/L10}

\bibitem[{{King} \& {Pounds}(2015)}]{2015ARA&A..53..115K}
{King}, A., \& {Pounds}, K. 2015, \araa, 53, 115, \dodoi{10.1146/annurev-astro-082214-122316}

\bibitem[{{Kormendy} \& {Ho}(2013)}]{2013ARA&A..51..511K}
{Kormendy}, J., \& {Ho}, L.~C. 2013, \araa, 51, 511, \dodoi{10.1146/annurev-astro-082708-101811}

\bibitem[{{Larkin} {et~al.}(2006){Larkin}, {Barczys}, {Krabbe}, {Adkins}, {Aliado}, {Amico}, {Brims}, {Campbell}, {Canfield}, {Gasaway}, {Honey}, {Iserlohe}, {Johnson}, {Kress}, {LaFreniere}, {Lyke}, {Magnone}, {Magnone}, {McElwain}, {Moon}, {Quirrenbach}, {Skulason}, {Song}, {Spencer}, {Weiss}, \& {Wright}}]{2006SPIE.6269E..1AL}
{Larkin}, J., {Barczys}, M., {Krabbe}, A., {et~al.} 2006, in Society of Photo-Optical Instrumentation Engineers (SPIE) Conference Series, Vol. 6269, Ground-based and Airborne Instrumentation for Astronomy, ed. I.~S. {McLean} \& M.~{Iye}, 62691A, \dodoi{10.1117/12.672061}

\bibitem[{{Li} {et~al.}(2023){Li}, {Tsai}, {Stern}, {Wu}, {Assef}, {Blain}, {D{\'\i}az-Santos}, {Eisenhardt}, {Griffith}, {Jarrett}, {Jun}, {Lake}, \& {Saade}}]{2023ApJ...958..162L}
{Li}, G., {Tsai}, C.-W., {Stern}, D., {et~al.} 2023, \apj, 958, 162, \dodoi{10.3847/1538-4357/ace25b}

\bibitem[{{Li} {et~al.}(2024){Li}, {Assef}, {Tsai}, {Wu}, {Eisenhardt}, {Stern}, {D{\'\i}az-Santos}, {Blain}, {Jun}, {Fern{\'a}ndez Aranda}, \& {Zewdie}}]{2024ApJ...971...40L}
{Li}, G., {Assef}, R.~J., {Tsai}, C.-W., {et~al.} 2024, \apj, 971, 40, \dodoi{10.3847/1538-4357/ad5317}

\bibitem[{{Lockhart} {et~al.}(2019){Lockhart}, {Do}, {Larkin}, {Boehle}, {Campbell}, {Chappell}, {Chu}, {Ciurlo}, {Cosens}, {Fitzgerald}, {Ghez}, {Lu}, {Lyke}, {Mieda}, {Rudy}, {Vayner}, {Walth}, \& {Wright}}]{2019AJ....157...75L}
{Lockhart}, K.~E., {Do}, T., {Larkin}, J.~E., {et~al.} 2019, \aj, 157, 75, \dodoi{10.3847/1538-3881/aaf64e}

\bibitem[{{Luo} {et~al.}(2022){Luo}, {Fan}, {Zou}, {Shen}, {Lin}, {Hu}, {Lin}, {Tao}, \& {Chen}}]{2022ApJ...935...80L}
{Luo}, Y., {Fan}, L., {Zou}, H., {et~al.} 2022, \apj, 935, 80, \dodoi{10.3847/1538-4357/ac8162}

\bibitem[{{Lyke} {et~al.}(2017){Lyke}, {Do}, {Boehle}, {Campbell}, {Chappell}, {Fitzgerald}, {Gasawy}, {Iserlohe}, {Krabbe}, {Larkin}, {Lockhart}, {Lu}, {Mieda}, {McElwain}, {Perrin}, {Rudy}, {Sitarski}, {Vayner}, {Walth}, {Weiss}, {Wizanski}, \& {Wright}}]{2017ascl.soft10021L}
{Lyke}, J., {Do}, T., {Boehle}, A., {et~al.} 2017, {OSIRIS Toolbox: OH-Suppressing InfraRed Imaging Spectrograph pipeline}, Astrophysics Source Code Library, record ascl:1710.021

\bibitem[{{Maiolino} {et~al.}(2012){Maiolino}, {Gallerani}, {Neri}, {Cicone}, {Ferrara}, {Genzel}, {Lutz}, {Sturm}, {Tacconi}, {Walter}, {Feruglio}, {Fiore}, \& {Piconcelli}}]{2012MNRAS.425L..66M}
{Maiolino}, R., {Gallerani}, S., {Neri}, R., {et~al.} 2012, \mnras, 425, L66, \dodoi{10.1111/j.1745-3933.2012.01303.x}

\bibitem[{{Martin} {et~al.}(2024){Martin}, {Blain}, {D{\'\i}az-Santos}, {Assef}, {Tsai}, {Jun}, {Eisenhardt}, {Wu}, {Vayner}, \& {Aranda}}]{2024MNRAS.534..978M}
{Martin}, L.~R., {Blain}, A.~W., {D{\'\i}az-Santos}, T., {et~al.} 2024, \mnras, 534, 978, \dodoi{10.1093/mnras/stae2147}

\bibitem[{{Mieda} {et~al.}(2014){Mieda}, {Wright}, {Larkin}, {Graham}, {Adkins}, {Lyke}, {Campbell}, {Maire}, {Do}, \& {Gordon}}]{2014PASP..126..250M}
{Mieda}, E., {Wright}, S.~A., {Larkin}, J.~E., {et~al.} 2014, \pasp, 126, 250, \dodoi{10.1086/675784}

\bibitem[{{Piconcelli} {et~al.}(2015){Piconcelli}, {Vignali}, {Bianchi}, {Zappacosta}, {Fritz}, {Lanzuisi}, {Miniutti}, {Bongiorno}, {Feruglio}, {Fiore}, \& {Maiolino}}]{2015A&A...574L...9P}
{Piconcelli}, E., {Vignali}, C., {Bianchi}, S., {et~al.} 2015, \aap, 574, L9, \dodoi{10.1051/0004-6361/201425324}

\bibitem[{{Rodrigo} \& {Solano}(2020)}]{2020sea..confE.182R}
{Rodrigo}, C., \& {Solano}, E. 2020, in XIV.0 Scientific Meeting (virtual) of the Spanish Astronomical Society, 182

\bibitem[{{Rodrigo} {et~al.}(2012){Rodrigo}, {Solano}, \& {Bayo}}]{2012ivoa.rept.1015R}
{Rodrigo}, C., {Solano}, E., \& {Bayo}, A. 2012, {SVO Filter Profile Service Version 1.0}, IVOA Working Draft 15 October 2012, \dodoi{10.5479/ADS/bib/2012ivoa.rept.1015R}

\bibitem[{{Rupke} \& {Veilleux}(2011)}]{2011ApJ...729L..27R}
{Rupke}, D. S.~N., \& {Veilleux}, S. 2011, \apjl, 729, L27, \dodoi{10.1088/2041-8205/729/2/L27}

\bibitem[{{Sanders} {et~al.}(1988){Sanders}, {Soifer}, {Elias}, {Madore}, {Matthews}, {Neugebauer}, \& {Scoville}}]{1988ApJ...325...74S}
{Sanders}, D.~B., {Soifer}, B.~T., {Elias}, J.~H., {et~al.} 1988, \apj, 325, 74, \dodoi{10.1086/165983}

\bibitem[{{Sanders} {et~al.}(2016){Sanders}, {Shapley}, {Kriek}, {Reddy}, {Freeman}, {Coil}, {Siana}, {Mobasher}, {Shivaei}, {Price}, \& {de Groot}}]{2016ApJ...816...23S}
{Sanders}, R.~L., {Shapley}, A.~E., {Kriek}, M., {et~al.} 2016, \apj, 816, 23, \dodoi{10.3847/0004-637X/816/1/23}

\bibitem[{{Santoro} {et~al.}(2020){Santoro}, {Tadhunter}, {Baron}, {Morganti}, \& {Holt}}]{2020A&A...644A..54S}
{Santoro}, F., {Tadhunter}, C., {Baron}, D., {Morganti}, R., \& {Holt}, J. 2020, \aap, 644, A54, \dodoi{10.1051/0004-6361/202039077}

\bibitem[{{Shen} {et~al.}(2019){Shen}, {Hall}, {Horne}, {Zhu}, {McGreer}, {Simm}, {Trump}, {Kinemuchi}, {Brandt}, {Green}, {Grier}, {Guo}, {Ho}, {Homayouni}, {Jiang}, {I-Hsiu Li}, {Morganson}, {Petitjean}, {Richards}, {Schneider}, {Starkey}, {Wang}, {Chambers}, {Kaiser}, {Kudritzki}, {Magnier}, \& {Waters}}]{2019ApJS..241...34S}
{Shen}, Y., {Hall}, P.~B., {Horne}, K., {et~al.} 2019, \apjs, 241, 34, \dodoi{10.3847/1538-4365/ab074f}

\bibitem[{{SSC And IRSA}(2020)}]{Spitzer2020}
{SSC And IRSA}. 2020, Spitzer Enhanced Imaging Products,  IPAC, \dodoi{10.26131/IRSA433}

\bibitem[{{Stern} {et~al.}(2014){Stern}, {Lansbury}, {Assef}, {Brandt}, {Alexander}, {Ballantyne}, {Balokovi{\'c}}, {Bauer}, {Benford}, {Blain}, {Boggs}, {Bridge}, {Brightman}, {Christensen}, {Comastri}, {Craig}, {Del Moro}, {Eisenhardt}, {Gandhi}, {Griffith}, {Hailey}, {Harrison}, {Hickox}, {Jarrett}, {Koss}, {Lake}, {LaMassa}, {Luo}, {Tsai}, {Urry}, {Walton}, {Wright}, {Wu}, {Yan}, \& {Zhang}}]{2014ApJ...794..102S}
{Stern}, D., {Lansbury}, G.~B., {Assef}, R.~J., {et~al.} 2014, \apj, 794, 102, \dodoi{10.1088/0004-637X/794/2/102}

\bibitem[{{Storey} \& {Zeippen}(2000)}]{2000MNRAS.312..813S}
{Storey}, P.~J., \& {Zeippen}, C.~J. 2000, \mnras, 312, 813, \dodoi{10.1046/j.1365-8711.2000.03184.x}

\bibitem[{{Sun} {et~al.}(2024){Sun}, {Fan}, {Han}, {Knudsen}, {Chen}, \& {Zhang}}]{2024ApJ...964...95S}
{Sun}, W., {Fan}, L., {Han}, Y., {et~al.} 2024, \apj, 964, 95, \dodoi{10.3847/1538-4357/ad22e3}

\bibitem[{{Tsai} {et~al.}(2015){Tsai}, {Eisenhardt}, {Wu}, {Stern}, {Assef}, {Blain}, {Bridge}, {Benford}, {Cutri}, {Griffith}, {Jarrett}, {Lonsdale}, {Masci}, {Moustakas}, {Petty}, {Sayers}, {Stanford}, {Wright}, {Yan}, {Leisawitz}, {Liu}, {Mainzer}, {McLean}, {Padgett}, {Skrutskie}, {Gelino}, {Beichman}, \& {Juneau}}]{2015ApJ...805...90T}
{Tsai}, C.-W., {Eisenhardt}, P. R.~M., {Wu}, J., {et~al.} 2015, \apj, 805, 90, \dodoi{10.1088/0004-637X/805/2/90}

\bibitem[{{Tsai} {et~al.}(2018){Tsai}, {Eisenhardt}, {Jun}, {Wu}, {Assef}, {Blain}, {D{\'\i}az-Santos}, {Jones}, {Stern}, {Wright}, \& {Yeh}}]{2018ApJ...868...15T}
{Tsai}, C.-W., {Eisenhardt}, P. R.~M., {Jun}, H.~D., {et~al.} 2018, \apj, 868, 15, \dodoi{10.3847/1538-4357/aae698}

\bibitem[{{Urrutia} {et~al.}(2008){Urrutia}, {Lacy}, \& {Becker}}]{2008ApJ...674...80U}
{Urrutia}, T., {Lacy}, M., \& {Becker}, R.~H. 2008, \apj, 674, 80, \dodoi{10.1086/523959}

\bibitem[{{Vayner} {et~al.}(2021){Vayner}, {Wright}, {Murray}, {Armus}, {Boehle}, {Cosens}, {Larkin}, {Mieda}, \& {Walth}}]{2021ApJ...919..122V}
{Vayner}, A., {Wright}, S.~A., {Murray}, N., {et~al.} 2021, \apj, 919, 122, \dodoi{10.3847/1538-4357/ac0f56}

\bibitem[{{Vayner} {et~al.}(2024){Vayner}, {D{\'\i}az-Santos}, {Eisenhardt}, {Stern}, {Armus}, {Angl{\'e}s-Alc{\'a}zar}, {Assef}, {Fern{\'a}ndez Aranda}, {Blain}, {Jun}, {Tsai}, {Roy}, {Brisbin}, {Ferkinhoff}, {Aravena}, {Gonz{\'a}lez-L{\'o}pez}, {Li}, {Liao}, {Shobhana}, {Wu}, \& {Zewdie}}]{2024arXiv241202862V}
{Vayner}, A., {D{\'\i}az-Santos}, T., {Eisenhardt}, P. R.~M., {et~al.} 2024, arXiv e-prints, arXiv:2412.02862, \dodoi{10.48550/arXiv.2412.02862}

\bibitem[{{Veilleux} {et~al.}(2005){Veilleux}, {Cecil}, \& {Bland-Hawthorn}}]{2005ARA&A..43..769V}
{Veilleux}, S., {Cecil}, G., \& {Bland-Hawthorn}, J. 2005, \araa, 43, 769, \dodoi{10.1146/annurev.astro.43.072103.150610}

\bibitem[{{Vito} {et~al.}(2018){Vito}, {Brandt}, {Stern}, {Assef}, {Chen}, {Brightman}, {Comastri}, {Eisenhardt}, {Garmire}, {Hickox}, {Lansbury}, {Tsai}, {Walton}, \& {Wu}}]{2018MNRAS.474.4528V}
{Vito}, F., {Brandt}, W.~N., {Stern}, D., {et~al.} 2018, \mnras, 474, 4528, \dodoi{10.1093/mnras/stx3120}

\bibitem[{{Wright} {et~al.}(2010){Wright}, {Eisenhardt}, {Mainzer}, {Ressler}, {Cutri}, {Jarrett}, {Kirkpatrick}, {Padgett}, {McMillan}, {Skrutskie}, {Stanford}, {Cohen}, {Walker}, {Mather}, {Leisawitz}, {Gautier}, {McLean}, {Benford}, {Lonsdale}, {Blain}, {Mendez}, {Irace}, {Duval}, {Liu}, {Royer}, {Heinrichsen}, {Howard}, {Shannon}, {Kendall}, {Walsh}, {Larsen}, {Cardon}, {Schick}, {Schwalm}, {Abid}, {Fabinsky}, {Naes}, \& {Tsai}}]{2010AJ....140.1868W}
{Wright}, E.~L., {Eisenhardt}, P. R.~M., {Mainzer}, A.~K., {et~al.} 2010, \aj, 140, 1868, \dodoi{10.1088/0004-6256/140/6/1868}

\bibitem[{Wright {et~al.}(2019)Wright, Eisenhardt, Mainzer, Ressler, Cutri, Jarrett, Kirkpatrick, Padgett, McMillan, Skrutskie, Stanford, Cohen, Walker, Mather, Leisawitz, III, McLean, Benford, Lonsdale, Blain, Mendez, Irace, Duval, Liu, Royer, Heinrichsen, Howard, Shannon, Kendall, Walsh, Larsen, Cardon, Schick, Schwalm, Abid, Fabinsky, Naes, \& Tsai}]{AllWISE2019}
Wright, E.~L., Eisenhardt, P. R.~M., Mainzer, A.~K., {et~al.} 2019, AllWISE Source Catalog,  IPAC, \dodoi{10.26131/IRSA1}

\bibitem[{{Wu} {et~al.}(2012){Wu}, {Tsai}, {Sayers}, {Benford}, {Bridge}, {Blain}, {Eisenhardt}, {Stern}, {Petty}, {Assef}, {Bussmann}, {Comerford}, {Cutri}, {Evans}, {Griffith}, {Jarrett}, {Lake}, {Lonsdale}, {Rho}, {Stanford}, {Weiner}, {Wright}, \& {Yan}}]{2012ApJ...756...96W}
{Wu}, J., {Tsai}, C.-W., {Sayers}, J., {et~al.} 2012, \apj, 756, 96, \dodoi{10.1088/0004-637X/756/1/96}

\bibitem[{{Wu} {et~al.}(2018){Wu}, {Jun}, {Assef}, {Tsai}, {Wright}, {Eisenhardt}, {Blain}, {Stern}, {D{\'\i}az-Santos}, {Denney}, {Hayden}, {Perlmutter}, {Aldering}, {Boone}, \& {Fagrelius}}]{2018ApJ...852...96W}
{Wu}, J., {Jun}, H.~D., {Assef}, R.~J., {et~al.} 2018, \apj, 852, 96, \dodoi{10.3847/1538-4357/aa9ff3}

\bibitem[{{Xu} \& {Wang}(2022)}]{2022ApJ...933..110X}
{Xu}, X., \& {Wang}, J. 2022, \apj, 933, 110, \dodoi{10.3847/1538-4357/ac7222}

\bibitem[{{Yuan} \& {Narayan}(2014)}]{2014ARA&A..52..529Y}
{Yuan}, F., \& {Narayan}, R. 2014, \araa, 52, 529, \dodoi{10.1146/annurev-astro-082812-141003}

\bibitem[{{Zewdie} {et~al.}(2023){Zewdie}, {Assef}, {Mazzucchelli}, {Aravena}, {Blain}, {D{\'\i}az-Santos}, {Eisenhardt}, {Jun}, {Stern}, {Tsai}, \& {Wu}}]{2023A&A...677A..54Z}
{Zewdie}, D., {Assef}, R.~J., {Mazzucchelli}, C., {et~al.} 2023, \aap, 677, A54, \dodoi{10.1051/0004-6361/202346695}

\bibitem[{{Zewdie} {et~al.}(2024){Zewdie}, {Assef}, {Lambert}, {Mazzucchelli}, {Ilani Loubser}, {Aravena}, {Gonz{\'a}lez-L{\'o}pez}, {Jun}, {Tsai}, {Stern}, {Li}, {Fern{\'a}ndez Aranda}, {D{\'\i}az-Santos}, {Eisenhardt}, {Vayner}, {Martin}, {Blain}, \& {Wu}}]{2024arXiv241204436Z}
{Zewdie}, D., {Assef}, R.~J., {Lambert}, T., {et~al.} 2024, arXiv e-prints, arXiv:2412.04436, \dodoi{10.48550/arXiv.2412.04436}

\bibitem[{{Zhan}(2011)}]{2011SSPMA..41.1441Z}
{Zhan}, H. 2011, Scientia Sinica Physica, Mechanica \& Astronomica, 41, 1441, \dodoi{10.1360/132011-961}

\end{thebibliography}
\bibliographystyle{aasjournal}

\end{document}